\documentclass[aps,pra,twocolumn,groupaddress]{revtex4-1}

\usepackage[dvips]{epsfig}
\usepackage{amsmath,amsfonts,amssymb}
\usepackage[sort&compress]{natbib}

\newcommand{\ket}[1]{\ensuremath{|#1\rangle}}

\begin{document}


\title{Prospects for Storage and Retrieval of a Quantum Dot Single Photon in an Ultracold $^{87}$Rb Ensemble}


\author{Matthew T. Rakher} \email{matthew.rakher@gmail.com}
\author{Richard J. Warburton}
\author{Philipp Treutlein} \email{philipp.treutlein@unibas.ch}
\affiliation{Departement Physik, Universit\"{a}t
Basel, Klingelbergstrasse 82, CH-4056 Basel, Switzerland}



\date{\today}

\begin{abstract}
Epitaxially grown quantum dots (QDs) are promising sources of non-classical states of light such as single photons and entangled photons.  However, in order for them to be used as a resource for long-distance quantum communication, distributed quantum computation, or linear optics quantum computing, these photons must be coupled efficiently to long-lived quantum memories as part of a quantum repeater network.  Here, we theoretically examine the prospects for efficient storage and retrieval of a QD-generated single photon with a 1~ns lifetime in a multi-level atomic system.  We calculate using an experimentally demonstrated optical depth of 150 that the storage (total) efficiency can exceed 46$\%$ (28$\%$) in a dense, ultracold ensemble of $^{87}$Rb atoms.  Furthermore, we find that the optimal control pulse required for storage and retrieval can be obtained using a diode laser and an electro-optic modulator rather than a mode-locked, pulsed laser source.  Increasing the optical depth, for example by using Bose-condensed ensembles or an optical cavity, can increase the efficiencies to near unity.  Aside from enabling a high-speed quantum network based on QDs, such an efficient optical interface between an atomic ensemble and a QD can also lead to entanglement between collective spin-wave excitations of atoms and the spin of an electron or hole confined in the QD.

\end{abstract}


\maketitle

\section{Introduction}\label{sec:intro}
Quantum communication and computation offer technological advances by performing information processing with quantum mechanics~\cite{ref:NielsenChuang,ref:Ladd_Nat10}.  In order to perform these tasks over large distances, photons are the obvious choice to carry quantum information.  However, transmission over large distances is dramatically hampered by attenuation in optical fibers.  Unlike classical fiber optic communication, amplifiers placed periodically along the transmission channel cannot be used to overcome this loss owing to the ``no-cloning" theorem~\cite{ref:Wootters_Nat_82}.  Fortunately, quantum repeaters have been proposed to resolve this issue by using entanglement shared between adjacent nodes and joint measurements to create entanglement between the start node and the terminal node~\cite{ref:Briegel_PRL_98, ref:Kimble_Nat08, ref:Sangouard_RMP_2011}.  A crucial element in this scheme is the ability to efficiently store and retrieve single photons using a quantum memory~\cite{ref:Duan_Lukin_Cirac_Zoller,ref:Lvovsky_Nphot_2009}.  To date, the essential ingredients of a photonic quantum memory have been demonstrated using ensembles of ultracold alkali atoms~\cite{ref:Chaneliere_Kuzmich,ref:Choi_Nat_2008,ref:ZhaoPan_NPhys_08,ref:ZhaoKuz_NPhys_09}, ensembles of ultracold atoms in cavities~\cite{ref:Simon_PRL_07}, warm vapors of alkali atoms~\cite{ref:Eisaman_Lukin,ref:Sherson_Nat_06,ref:Reim_NatPhot10}, ultracold single atoms in cavities~\cite{ref:Specht_Nat_11}, and solid-state systems composed of rare-earth dopants in crystals~\cite{ref:Clausen_Nat11, ref:Saglamyurek_Nat11}.  Specifically, retrieval efficiencies as high as 73$\%$ and storage times as long as 3.2~ms have been simultaneously demonstrated using ultracold atoms~\cite{ref:Bao_Nphys_2012}.  The combination of large optical depths and long ground state hyperfine coherence make ultracold atomic ensembles an attractive platform for optical quantum memories.

While there are quantum memory schemes which use probabilistically-generated spin waves from spontaneous Raman scattering in atomic ensembles, it has been shown that schemes based on fast (rates approaching the GHz scale) single photon or entangled photon sources can provide better performance~\cite{ref:Sangouard_PRA_2007, ref:Sangouard_RMP_2011}.  Thus, one would ideally like a source of quantum light states that is on-demand and bright, meaning it can produce these photons on fast timescales (broadband) in a triggered fashion.  In addition, these broadband photons should be indistinguishable so they bunch perfectly on a beamsplitter~\cite{ref:Knill_Nat_01}.  A promising source of such on-demand, non-classical light states are epitaxially grown, GaAs-based semiconductor quantum dots (QDs)~\cite{ref:Shields_NPhot}.  The short spontaneous emission lifetime of a QD, more than an order of magnitude shorter than that of an alkali atom, and the fact that it is embedded in a robust optoelectronic material make it an attractive candidate as a quantum light source.  As an individual two-level system, the QD naturally emits one photon when it is excited and collection of this photon can be very efficient by proper design of the surrounding dielectric~\cite{ref:Strauf_NPhot, ref:Claudon,ref:Davanco_APL_11}.  Single photon count rates can in principle approach 1 GHz, with higher rates possible by taking advantage of a Purcell enhancement.  In addition, QDs have been shown to emit polarization entangled photons by means of a cascaded decay~\cite{ref:Stevenson_Nat_06,ref:Muller_PRL_09}.  Finally, QD-generated photons can have a high degree of indistinguishability~\cite{ref:Ates_PRL09,ref:He_Nnano_2013}. Aside from these optical properties, the internal spin states of charged quantum dots have received considerable attention for quantum information processing~\cite{ref:Atature_Sci_06,ref:Gerardot_Nat,ref:deGreve_NPhys_11}.  Spin coherence times as along as 1~$\mu$s have been measured~\cite{ref:Brunner_Sci_09} and because the spin state can be entangled with the polarization of an emitted photon~\cite{ref:Simon_PRB_07,ref:DeGreve_Nat_12,ref:Gao_Nat_12}, charged quantum dots are a natural candidate for a solid-state qubit that interacts strongly with light.

While QDs are promising sources of single or entangled photons, they must be coupled to a high quality quantum memory in order to be a viable source for long distance quantum information processing.  Here, we investigate the storage and retrieval of a broadband, QD-generated single photon with a 1 ns lifetime in an ultracold, dense ensemble of $^{87}$Rb atoms.  Taking previously measured experimental parameters, we find that the total efficiency ($\eta_{tot} = \eta_{s} \times \eta_{r}$, where $\eta_s$ and $\eta_r$ are the storage and retrieval efficiencies) can exceed 28$\%$ for storage and backwards retrieval of a photon with a 1~ns lifetime in a $^{87}$Rb ensemble with an on-resonance optical depth of 150.  Because the bandwidth of the QD photon can approach the excited-state hyperfine splitting in $^{87}$Rb, this result was obtained by extending the $\Lambda$-system theory of Gorshkov \textit{et al} \cite{ref:Gorshkov_PRL07,ref:Gorshkov_PRA2} to a four-level system.  Using the gradient ascent approach outlined in Ref.~\cite{ref:Gorshkov_PRA4}, we determine the maximum efficiency as well as the optimal control pulse for photon storage.  We find that the control pulse can be easily generated using a 12~mW laser diode in contrast to the broadband, Raman-based scheme in Ref.~\cite{ref:Reim_NatPhot10} where a mode-locked laser is required.  Finally, we consider the effects of excess dephasing and spectral wandering of the QD optical transition and show that the memory efficiency remains fairly robust.  These results show that with existing technology, QD-generated photons can be reliably interfaced with ultracold atomic ensembles, paving the way for their future use in quantum information as well as opening interesting avenues in the study of hybrid quantum systems.

This paper is arranged as follows: In Sec.~\ref{sec:3L}, we briefly review the physics of photon storage and retrieval in a $\Lambda$-system following the treatment of Gorshkov \textit{et al} \cite{ref:Gorshkov_PRL07,ref:Gorshkov_PRA2}.  In Sec.~\ref{sec:4L}, we extend this treatment to a four-level atom, which is relevant for broadband photon storage in atomic systems with non-negligible hyperfine structure in the excited state.  Section \ref{sec:QD} implements the model to study storage and retrieval of a broadband single photon emitted by a quantum dot using a $^{87}$Rb ensemble.  In Sec.~\ref{sec:dephase}, deleterious effects resulting from imperfect indistinguishability of the quantum dot photon are discussed.  In sec.~\ref{sec:highOD} we investigate how the efficiencies increase for very high optical depth.  The work is concluded and summarized in Sec.~\ref{sec:conc} and details of the calculations and numerical implementation are discussed in the Appendices A and B.  Appendix C discusses the role of four-wave mixing (FWM) in the storage process.
\section{Review of Photon Storage in a $\Lambda$-system}\label{sec:3L}
Photon storage in a three level, $\Lambda$-type system has been the subject of many articles and reviews~\cite{ref:Sangouard_RMP_2011}.  Here, we briefly touch on the main points and restrict the discussion to those schemes described by the theory of Gorshkov \textit{et al}, namely those based on Electromagnetically Induced Transparency (EIT), off-resonant Raman, or photon echo interactions~\cite{ref:Gorshkov_PRL07,ref:Gorshkov_PRA2}.  In that work, these three schemes were shown to yield similar storage efficiencies for the same optical depth of the $\Lambda$ medium and are in this sense equivalent.  However, there are tradeoffs in the actual physical implementation for these schemes which will be discussed later.
\begin{figure}[h]
\centering
\includegraphics[width=\columnwidth, trim = 0 65 0 0, clip=true]{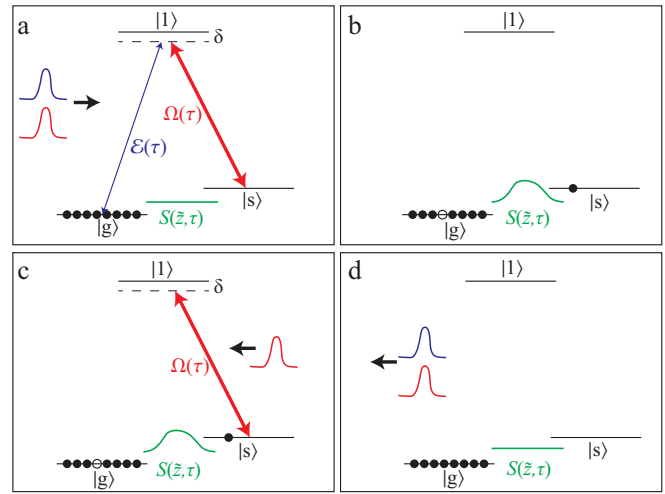}
\caption{Simple schematic of photon storage and retrieval in a $\Lambda$ system. (a) A quantum field $\mathcal{E}(\tau)$ and a classical control field $\Omega(\tau)$ with a common excited state detuning $\delta$ impinge on an ensemble of $\Lambda$-type atomic systems from the left.  All atoms are initialized into the state $\ket{g}$.  (b) The quantum field is transformed and stored into a ground state coherence (spin wave) $S(\tilde{z},\tau)$. (c) The spin wave is converted back into a photon by exciting the system with another classical control field from the right. (d) The single photon exits the ensemble propagating to the left.} \label{fig:fig1}
\end{figure}

The basic picture of photon storage in such a system is shown in Fig.~\ref{fig:fig1}.  Each atom in the ensemble is composed of two ground states $\ket{g}$ and $\ket{s}$ that can be optically coupled to the excited state $\ket{1}$.  These two transitions can be individually addressed using a sufficient ground state splitting or by selection rules so as to avoid cross-coupling.  The storage procedure starts with all of the atoms initialized into $\ket{g}$.  Then a quantum field $\mathcal{E}(\tau)$ which is to be stored, addresses the $\ket{g}-\ket{1}$ transition while a classical control field $\Omega(\tau)$ addresses the $\ket{s}-\ket{1}$ transition (see Fig.~\ref{fig:fig1}(a)).  The control pulse facilitates the transfer of the quantum field to a spatially-dependent coherence of the two ground states (often called a ``spin wave") $S(\tilde{z},\tau)$ as depicted in Fig.~\ref{fig:fig1}(b).  The exact physical mechanism that accomplishes this task depends on the scheme.  In EIT storage, the control pulse dynamically creates a transparency window in the absorption profile of the $\ket{g}-\ket{1}$ transition while adiabatically reducing the group velocity of the quantum field to 0~\cite{ref:Lukin_RMP_03}.  In Raman storage, the control pulse enables the quantum field to be absorbed into $\ket{s}$ by a two-photon Raman transition~\cite{ref:Nunn_PRA_07}.  For photon echo storage, the quantum field promotes an atom to the excited state $\ket{1}$ and the control pulse transfers the excitation to $\ket{s}$ by performing a fast $\pi$ pulse.  In all cases, the quantum field transfers one atom from $\ket{g}$ to $\ket{s}$, however there is no knowledge of which atom.  Thus the excitation is coherently distributed over the entire ensemble and it is this collective behavior that makes the process efficient.

The quantum field is now stored in this collective ground state coherence and is therefore sensitive to decoherence processes.  These processes, which can include magnetic field fluctuations or atomic motion, set the limit for the spin wave coherence time, and hence, the duration the photon can be stored.  Notably, the spin wave is more robust against decoherence than many other multi-particle entangled states~\cite{ref:Lukin_RMP_03}.  The spin wave can be re-converted into a photon by using another control pulse (see Fig.~\ref{fig:fig1}(c)-(d)).  Gorshkov \textit{et al} showed that the optimal retrieval is simply the time-reverse of storage; so-called ``backwards retrieval"~\cite{ref:Gorshkov_PRL07}.  This process creates a photon propagating in the opposite direction to its initial propagation.  Forwards retrieval is also possible, but will not be treated here.

\begin{figure}[h]
\centering
\includegraphics[width=0.65\columnwidth, trim = 0 65 0 0, clip=true]{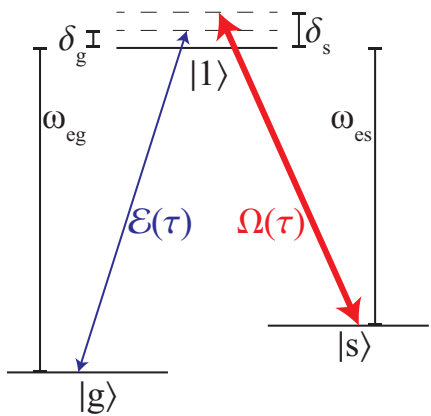}
\caption{Diagram of the level structure considered for photon storage.  The quantum field $\mathcal{E}(\tau)$ oscillates at frequency $\nu_{eg} = \omega_{eg}+\delta_g$ while the classical control field $\Omega(\tau)$ oscillates at frequency $\nu_{es} = \omega_{es}+\delta_s$.} \label{fig:fig2}
\end{figure}
For all of these storage schemes, the dynamics of the interaction between the two optical fields and the atoms is described by treating the signal field on the $\ket{g}-\ket{1}$ transition quantum mechanically while treating the $\ket{s}-\ket{1}$ control field semiclassically.  The atomic level structure with relevant energy scales for photon storage in a three-level $\Lambda$ system is shown in Fig.~\ref{fig:fig2}.  The ensemble of $\Lambda$ systems is composed of $N$ atoms distributed over a volume of length $L$ and cross-sectional area $A$ with linear density $n(z)$.  In this analysis, we ignore the motion of the atoms and thereby restrict the discussion to ultracold ensembles.  As was shown in Ref.~\cite{ref:Gorshkov_PRA2}, the important parameter governing the efficiency of the storage process is the optical depth, defined as
\begin{equation}
d = \frac{g^2 N L}{\gamma c},
\end{equation}
where $g$ is the single-photon coupling constant for the transition and $\gamma$ is the decay of the coherence of the excited state (for a purely radiative decay with no additional dephasing $2\gamma=\Gamma$, the spontaneous emission rate).  Note that the $d$ used here and in Ref.~\cite{ref:Gorshkov_PRA2} is equal to half of the optical depth as usually defined ($2d = d_{std} = \sigma \rho L$ where $\sigma$ is the resonant absorption cross-section of a single atom and $\rho$ is the number density).  The classical control field is described by a Rabi frequency envelope $\Omega(z,t)=\Omega(z-t/c)$ centered at frequency $\nu_{es} = \omega_{es}+\delta_s$.  As also noted in App.~A, this Rabi frequency is defined to be 1/2 of the usually defined Rabi frequency.  The quantum field is described by a slowly varying envelope operator $\mathcal{\hat{E}}(z,t)$ centered at frequency $\nu_{eg} = \omega_{eg}+\delta_g$.  In addition to $\mathcal{\hat{E}}(z,t)$, there are two other operators required to describe the dynamics: the polarization operator $\hat{P}_1(z,t) = \sqrt{N}\hat{\sigma}_{g1}(z,t)$ and the spin-wave operator $\hat{S}(z,t) = \sqrt{N}\hat{\sigma}_{gs}(z,t)$ where $\hat{\sigma}_{\alpha \beta}(z,t)$ are slowly-varying collective atomic operators defined in App.~A.  It was shown in Ref.~\cite{ref:Gorshkov_PRA1} that in order to determine normally-ordered quantities such as efficiencies, one can neglect quantum noise operators and treat all dynamical variables as complex numbers.

The equations of motion governing $\mathcal{E}$, $P_1$, and $S$ are derived by calculating the Heisenberg equation of motion for all dynamical variables using the dipole and rotating wave approximations.  Then, two further approximations are made related to the fact that the quantum field is weak.  First, it is assumed that almost all of the atoms remain in the ground state $\ket{g}$ during the whole process.  Second, only terms to linear order in $\mathcal{E}$ are retained.  Under these approximations (see Ref.~\cite{ref:Gorshkov_PRA2} and App.~A for details), the following equations of motion are obtained:

\begin{align}
\partial_{\tilde{z}} \mathcal{E} &= i \tilde{\mu}_{1g}\sqrt{d \gamma}P_1 \label{eqn:3LEoME}\\
\partial_{\tau}P_1 &= \left(i \delta_g-\gamma \right) P_1+i \tilde{\mu}_{1s} \Omega(\tau) S +i \tilde{\mu}_{1g} \sqrt{d \gamma} \mathcal{E} \label{eqn:3LEoMP1}\\
\partial_{\tau}S &= i \left(\delta_g - \delta_s \right) S + i \tilde{\mu}_{1s}\Omega^*(\tau)P_1. \label{eqn:3LEoMS}
\end{align}
These equations of motion use a coordinate system $(\tilde{z},\tau)$, where $\tilde{z} = (1/N) \int_0^z dz'\, n(z')$ is a dimensionless length parameter ($\tilde{z} \in [0,1] $) and $\tau=t-z/c$ is the time in a co-moving reference frame.  Compared to the results of Ref.~\cite{ref:Gorshkov_PRA2}, Eqs.~\ref{eqn:3LEoME}-\ref{eqn:3LEoMS} also include relative dipole moments $\tilde{\mu}_{\alpha \beta}$.  These are defined as dipole moments relative to that of the two-level, cycling transition $\mu_{cyc}$ where a measurement of the optical depth would take place.  This sets the natural scale of the atom's dipole strength and allows for easy comparison between different physical implementations of levels $\ket{g}$, $\ket{s}$, and $\ket{1}$ (see App.~A for details).

For storage, we would like to take a quantum field with initial envelope $\mathcal{E}_{\textrm{in}}(\tau)$ (non-zero on the interval $\tau \in [0,T]$) and map it into a spin wave $S(\tilde{z},T)$ using a classical pulse $\Omega(\tau)$. The boundary conditions for the dynamical variables are $\mathcal{E}(0,\tau)=\mathcal{E}_{\textrm{in}}(\tau)$ and $S(\tilde{z},0)=P_1(\tilde{z},0)=0$.  We then want to compute the efficiency of this mapping once $\mathcal{E}_{\textrm{in}} = 0$ at $\tau = T$.  If the envelope of the initial quantum field is normalized ($\int_0^T d\tau \left|\mathcal{E}_{\textrm{in}}(\tau) \right|^2=1$), the storage efficiency $\eta_s$ is given by~\cite{ref:Gorshkov_PRA2}
\begin{equation}
\eta_s = \int_0^1 d\tilde{z}\left|S(\tilde{z},T)\right|^2 \label{eqn:etastor}.
\end{equation}

For a given input field and optical depth of the medium, one would like to determine the optimal classical pulse shape so as to maximize the storage efficiency.  In Ref.~\cite{ref:Gorshkov_PRA2} it was shown that in the adiabatic limit, $Td\gamma \gg 1$, an analytic solution for $\Omega(\tau)$ could be found and that the efficiency scaled as $\eta_s \propto d/(1+d)$.  For broadband photon storage, the adiabatic limit is not necessarily met and the optimal $\Omega(\tau)$ must be found numerically.  Gorshkov \textit{et al.} used a gradient ascent algorithm in Ref.~\cite{ref:Gorshkov_PRA4} to numerically optimize $\eta_s$ and found that the results matched the analytical solution in the adiabatic limit.  In addition, it was shown that photons beyond the adiabatic regime ($1/T \approx d \gamma$) could also be stored efficiently by using this optimization technique. This is the approach we will use in the following analysis and now briefly review.  Because we take a numerical approach, we now restrict the discussion to near-resonant storage schemes and exclude off-resonant Raman based storage which would require a much larger computational domain so as to capture the rapidly varying detuning ($\delta_g,\delta_s \gg \sqrt{d \gamma/T_1}$).  Gorshkov \textit{et al} mentioned in Ref.~\cite{ref:Gorshkov_PRA4} how one can obtain the optimal control pulse for Raman-based storage from the optimal control pulse found for resonant storage, but we do not pursue this here.  We do, however, note that the control pulse required for Raman storage can be several orders of magnitude more intense than that required for resonant schemes~\cite{ref:Reim_NatPhot10} for the same optical depth $d$ and hence the same efficiency.

A gradient ascent algorithm simply starts with a trial solution and proceeds to the optimal solution by moving along the gradient of the quantity to be maximized.  At each step in the algorithm, the gradient is determined and the control pulse is updated.  Mathematically, this replacement rule is
\begin{equation}
\Omega(\tau) \rightarrow \Omega(\tau)+\lambda \frac{\delta J}{\delta \Omega(\tau)},
\end{equation}
where $\lambda$ is the step size parameter and $J$ is the quantity to be maximized.  In the case of photon storage, we want to maximize the storage efficiency subject to the constraint that all of the dynamical variables fulfill the equations of motion (Eq.~\ref{eqn:3LEoME}-\ref{eqn:3LEoMS}).  Thus, $J$ takes the form
\begin{align}
J &=\int_0^1 d\tilde{z}\,S(\tilde{z},T)S^*(\tilde{z},T) \nonumber \\
&+\int_0^T d\tau \int_0^1 d\tilde{z}\,\left\{ \bar{\mathcal{E}}^* \left[ -\partial_{\tilde{z}} \mathcal{E} + i \tilde{\mu}_{1g}\sqrt{d \gamma}P_1 \right] +c.c.\right\} \nonumber \\
&+\int_0^T d\tau \int_0^1 d\tilde{z}\, \Big\{ \bar{P}_1^*\Big[ \left(-\partial_{\tau}+i \delta_g-\gamma \right) P_1+i \tilde{\mu}_{1s} \Omega(\tau) S  \nonumber \\
&+ i \tilde{\mu}_{1g} \sqrt{d \gamma} \mathcal{E} \Big] +c.c. \Big\} \nonumber \\
&+\int_0^T d\tau \int_0^1 d\tilde{z}\,\Big\{ \bar{S}^*\Big[-\partial_{\tau}S+ i \left(\delta_g - \delta_s \right) S \nonumber \\
&+ i \tilde{\mu}_{1s}\Omega^*(\tau)P_1  \Big] +c.c. \Big\}, \label{eqn:3LJ}
\end{align}
where the first term is $\eta_s$ and Lagrange multipliers $\bar{\mathcal{E}}$, $\bar{P}_1$, and $\bar{S}$ have been introduced to include the equations of motion.  The maximum storage efficiency is found when $J$ is stationary with respect to variations in all dynamical variables and $\Omega(\tau)$.  Requiring stationarity with respect to variations in $\mathcal{E}$, $P_1$, and $S$ results in equations of motion and boundary conditions for the Lagrange multipliers $\bar{\mathcal{E}}$, $\bar{P}_1$, and $\bar{S}$.  The equations of motion are
\begin{align}
\partial_{\tilde{z}} \bar{\mathcal{E}} &= i \tilde{\mu}_{1g}\sqrt{d \gamma}\bar{P}_1 \label{eqn:3LEoMbE}\\
\partial_{\tau}\bar{P}_1 &= \left(i \delta_g+\gamma \right) \bar{P}_1+i \tilde{\mu}_{1s} \Omega(\tau) \bar{S} +i \tilde{\mu}_{1g} \sqrt{d \gamma} \bar{\mathcal{E}} \label{eqn:3LEoMbP1}\\
\partial_{\tau}\bar{S} &= i \left(\delta_g - \delta_s \right) \bar{S} + i \tilde{\mu}_{1s}\Omega^*(\tau)\bar{P}_1 \label{eqn:3LEoMbS}
\end{align}
with boundary conditions $\bar{S}(\tilde{z},T) = S(\tilde{z},T)$ and $\bar{\mathcal{E}}(1,\tau)=\bar{P}_1(\tilde{z},T)=0$.  As pointed out in Ref.~\cite{ref:Gorshkov_PRA4}, these are exactly the equations of motion and boundary conditions for backwards retrieval.  Thus, solving these equations will yield the retrieval efficiency, $\eta_r$, and the total efficiency $\eta_{tot} = \eta_s \eta_r$ for storage followed by backwards retrieval.  Explicitly, these quantities can be determined using
\begin{align}
\eta_{r} &= \frac{\int_0^T d\tau \, |\mathcal{E}_{\textrm{out}}(\tau)|^2}{\int_0^1 d\tilde{z}\left|S(\tilde{z},T)\right|^2} \\
\eta_{tot} &= \int_0^T d\tau \, |\mathcal{E}_{\textrm{out}}(\tau)|^2
\end{align}
where $\mathcal{E}_{\textrm{out}}(\tau) = \bar{\mathcal{E}}(0,\tau)$ is the output quantum field.

The variation of $J$ with respect to variations in $\Omega(\tau)$ (the gradient) can be identified from Eq.~\ref{eqn:3LJ} as
\begin{equation}
\frac{\delta J}{\delta \Omega(\tau)} = -2 \tilde{\mu}_{1s} \int_0^1 d\tilde{z}\, \mathrm{Im}{\left[\bar{S}^*P_1-\bar{P}_1 S^* \right]}. \label{eqn:3Lgrad}
\end{equation}
Thus, the prescription for obtaining the maximum storage efficiency and optimal control pulse for a given input quantum field and optical depth is as follows.  First, take a trial control pulse and solve the equations of motion Eq.~\ref{eqn:3LEoME}-\ref{eqn:3LEoMS} with the storage boundary conditions to obtain $P_1(\tilde{z},\tau)$ and $S(\tilde{z},\tau)$.  Then, solve Eq.~\ref{eqn:3LEoMbE}-\ref{eqn:3LEoMbS} using the boundary conditions for backwards retrieval (note that these equations run backwards in time and space) to obtain $\bar{P}_1(\tilde{z},\tau)$ and $\bar{S}(\tilde{z},\tau)$.  Now the gradient can be determined from Eq.~\ref{eqn:3Lgrad} and the control pulse can be updated using the replacement rule.  This process is then repeated until the desired tolerance of the storage (or total) efficiency is obtained.

\begin{figure}[h]
\centering
\includegraphics[width=0.95\columnwidth, clip=true]{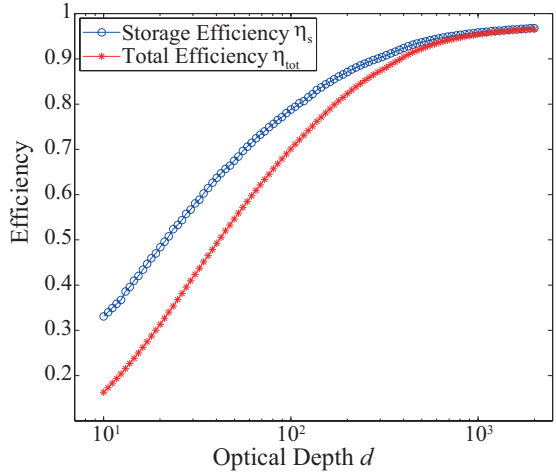}
\caption{(Color online) Results of the numerical calculation using gradient ascent for storage and retrieval of a quantum dot photon in an idealized Rb gas.  The storage (total) efficiency is plotted as a function of the optical depth $d$ in blue (red).  Solid curves are guides to the eye.} \label{fig:fig3}
\end{figure}
As an example we consider the case of an idealized Rb gas ($\tilde{\mu}_{1g} = \tilde{\mu}_{1s}=1$, $\gamma = 2\pi\times3.035$ MHz) storing a quantum-dot photon with $\mathcal{E}_{\textrm{in}}(\tau) = \Theta(\tau)\exp(-\tau/2 T_1)/\sqrt{T_1}$ where $\Theta(\tau)$ is the Heaviside step function and $T_1$ is the spontaneous emission lifetime (taken to be 1~ns).  We take both fields to be exactly on resonance $\delta_g=\delta_s=0$.  As a function of $d$, the storage and total efficiencies are determined by performing gradient ascent for each value of $d$.  Details and limitations of the numerical implementation can be found in Appendix~\ref{app:Num}.  The results of the numerical calculation are shown in Fig.~\ref{fig:fig3} where the storage (blue) and total (red) efficiencies are plotted as a function of $d$.  Both curves reach an asymptotic value of $\approx$96.0$\%$ for $d\approx10^3$ limited by the finite computational domain and the instantaneous rise of the quantum field, which is non-physical but chosen for calculational simplicity (see App.~\ref{app:Num}).

\begin{figure}[!h]
\centering
\includegraphics[width=0.95\columnwidth, clip=true]{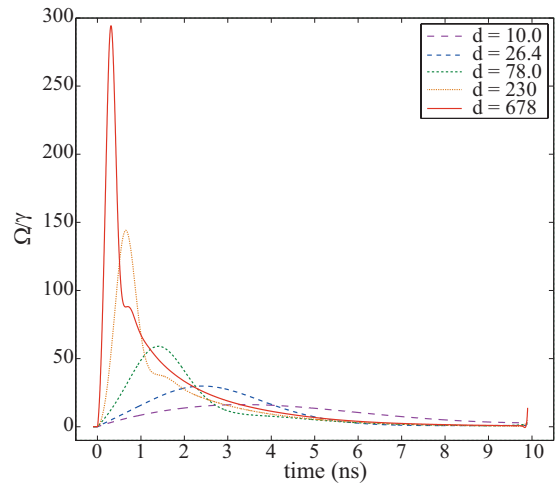}
\caption{(Color online) Optimal control pulses for quantum dot photon storage with $d=\{10.0,26.4,78.0,230,678\}$ determined using gradient ascent.  The Rabi frequencies are scaled by $\gamma = 2\pi\times3.035$ MHz.}   \label{fig:fig4}
\end{figure}
In addition to determining the efficiencies, it is instructive to see how the optimal control pulse $\Omega(\tau)$ changes as a function of $d$.  Figure~\ref{fig:fig4} shows the optimal control pulse for a few selected values of $d$ and it can be seen that the required peak Rabi frequency $\Omega_m$ increases with $d$.  For these parameters, we find that $\Omega_m \propto d^{0.67}$.  Furthermore, $\Omega(\tau)$ starts to become much more heavily weighted at the start of the pulse.  Both of these observations can be understood as the optimal storage scheme slowly changes from a photon echo type storage towards an EIT-based storage as the adiabatic limit is approached.  Lastly, the widths of these control pulses are not ultrafast, i.e. they do not require a mode-locked laser but rather are achievable with more flexible techniques such as direct intensity modulation.  The combination of relatively high total efficiencies and realistic control pulses make it seem that storage of a quantum dot photon in an atomic gas like Rb is certainly feasible with current technology.

Importantly, the peak Rabi frequency $\Omega_m$ of the control pulse surpasses 50$\gamma$ even for relatively low $d$.  For such large values of $\Omega_m$, the more complicated level structure of the storage medium will begin to play a significant role in the dynamics.  For instance, the excited-state splittings of the commonly-used $D_2$ transitions of $^{87}$Rb are around 100 MHz ($\approx 30\gamma$).  As a result, the three-level $\Lambda$ treatment is insufficient to describe broadband, on-resonance storage and more levels must be taken into account.

\section{Photon Storage in a Four-level System}\label{sec:4L}

\begin{figure}[h]
\centering
\includegraphics[width=0.65\columnwidth, trim = 0 65 0 0, clip=true]{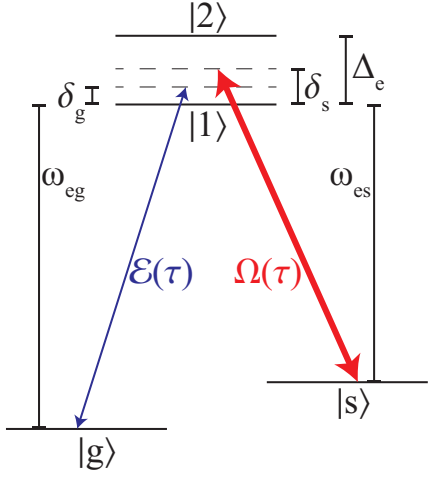}
\caption{Diagram of the level structure considered for photon storage in a four-level system.  The structure is the same as Fig.~\ref{fig:fig2} except an additional excited state $\ket{2}$ is included at an energy $\Delta_e$ above $\ket{1}$.} \label{fig:fig5}
\end{figure}

We now extend the treatment of Sec.~\ref{sec:3L} to include another excited state that can couple to both the control field and the input quantum field as shown in Fig.~\ref{fig:fig5}.  Because the maximum Rabi frequency of the control pulse can exceed the excited state level splitting $\Delta_e$, an atom can go from $\ket{g}$ to $\ket{s}$ via $\ket{1}$ or $\ket{2}$.  Depending on the dipole matrix elements $\tilde{\mu}_{\alpha \beta}$, these paths can interfere constructively or destructively.


Following the same procedure as in Sec.~\ref{sec:3L} and adding another polarization $\hat{P}_2(z,t) = \sqrt{N}\hat{\sigma}_{g2}(z,t)$ results in the following equations of motion (see App.~A for details):

\begin{align}
\partial_{\tilde{z}} \mathcal{E} &= i\sqrt{d \gamma}\left[\tilde{\mu}_{1g} P_1 + \tilde{\mu}_{2g} P_2\right] \label{eqn:4LEoME}\\
\partial_{\tau}P_1 &= \left(i \delta_g-\gamma \right) P_1+i \tilde{\mu}_{1s} \Omega(\tau) S +i \tilde{\mu}_{1g} \sqrt{d \gamma} \mathcal{E} \label{eqn:4LEoMP1}\\
\partial_{\tau}P_2 &= \left(i\delta_g -i\Delta_e-\gamma\right) P_2 + i \tilde{\mu}_{2s} \Omega(\tau) S \nonumber \\
&+ i \tilde{\mu}_{2g} \sqrt{d \gamma} \mathcal{E} \label{eqn:4LEoMP2}\\
\partial_{\tau}S &= i \left(\delta_g - \delta_s \right) S + i\Omega^*(\tau)\left[ \tilde{\mu}_{1s}P_1 + \tilde{\mu}_{2s}P_2 \right] \label{eqn:4LEoMS}
\end{align}
where new relative dipole moments $\tilde{\mu}_{2g}$ and $\tilde{\mu}_{2s}$ have been introduced to represent the couplings to the additional excited state.  As the spin wave $S(\tilde{z},\tau)$ and the field $\mathcal{E}(\tilde{z},\tau)$ are unchanged, the definitions of the storage, retrieval, and total efficiencies remain the same.  Furthermore, the boundary conditions for storage are the same as before except $P_2(\tilde{z},0)=0$ is added.

We can also find the four-level version of $J$
\begin{align}
J &=\int_0^1 d\tilde{z}\,S(\tilde{z},T)S^*(\tilde{z},T)+\int_0^T d\tau \int_0^1 d\tilde{z}\,\Big\{ \bar{\mathcal{E}}^* \Big[ -\partial_{\tilde{z}} \mathcal{E} \nonumber\\
&+ i \sqrt{d \gamma}\left(\tilde{\mu}_{1g} P_1 +\tilde{\mu}_{2g} P_2\right)\Big] +c.c.\Big\} \nonumber \\
&+\int_0^T d\tau \int_0^1 d\tilde{z}\, \Big\{ \bar{P}_1^*\Big[(-\partial_{\tau}+i \delta_g-\gamma) P_1+i \tilde{\mu}_{1s} \Omega(\tau) S  \nonumber \\
&+ i \tilde{\mu}_{1g} \sqrt{d \gamma} \mathcal{E} \Big] +c.c. \Big\} \nonumber \\
&+\int_0^T d\tau \int_0^1 d\tilde{z}\, \Big\{ \bar{P}_2^*\Big[ \left(-\partial_{\tau}+i\delta_g-i\Delta_e-\gamma  \right) P_2  \nonumber \\
&+i \tilde{\mu}_{2s} \Omega(\tau) S+ i \tilde{\mu}_{2g} \sqrt{d \gamma} \mathcal{E} \Big] +c.c. \Big\} \nonumber \\
&+\int_0^T d\tau \int_0^1 d\tilde{z}\,\Big\{ \bar{S}^*\Big[-\partial_{\tau}S+ i \left(\delta_g - \delta_s \right) S \nonumber \\
&+ i\Omega^*(\tau)\left( \tilde{\mu}_{1s}P_1+\tilde{\mu}_{2s}P_2  \right) \Big] +c.c. \Big\}, \label{eqn:4LJ}
\end{align}
in order to obtain the optimal control and efficiencies.  By requiring that $J$ is stationary with respect to variations in $\mathcal{E}$, $P_1$, $P_2$, and $S$, we find the equations of motion
\begin{align}
\partial_{\tilde{z}} \bar{\mathcal{E}} &= i\sqrt{d \gamma}\left( \tilde{\mu}_{1g}\bar{P}_1+\tilde{\mu}_{2g}\bar{P}_2 \right)\label{eqn:4LEoMbE}\\
\partial_{\tau}\bar{P}_1 &= \left(i \delta_g+\gamma \right) \bar{P}_1+i \tilde{\mu}_{1s} \Omega(\tau) \bar{S} +i \tilde{\mu}_{1g} \sqrt{d \gamma} \bar{\mathcal{E}} \label{eqn:4LEoMbP1}\\
\partial_{\tau}\bar{P}_2 &= \left(i \delta_g-i\Delta_e+\gamma \right) \bar{P}_2+i \tilde{\mu}_{2s} \Omega(\tau) \bar{S} +i \tilde{\mu}_{2g} \sqrt{d \gamma} \bar{\mathcal{E}} \label{eqn:4LEoMbP2}\\
\partial_{\tau}\bar{S} &= i \left(\delta_g - \delta_s \right) \bar{S} + i\Omega^*(\tau)\left( \tilde{\mu}_{1s}\bar{P}_1+\tilde{\mu}_{2s}\bar{P}_2 \right)\label{eqn:4LEoMbS}
\end{align}
for the Lagrange multipliers and their boundary conditions ($\bar{S}(\tilde{z},T) = S(\tilde{z},T)$ and $\bar{\mathcal{E}}(1,\tau)=\bar{P}_1(\tilde{z},T)=\bar{P}_2(\tilde{z},T)=0$).  The gradient along $\Omega(\tau)$ is now modified to
\begin{align}
\frac{\delta J}{\delta \Omega(\tau)} &= -2 \int_0^1 d\tilde{z}\, \mathrm{Im}\Big[\bar{S}^*\left(\tilde{\mu}_{1s} P_1+\tilde{\mu}_{2s} P_2 \right) \nonumber \\ &-\left(\tilde{\mu}_{1s}\bar{P}_1+\tilde{\mu}_{2s}\bar{P}_2 \right) S^* \Big]. \label{eqn:4Lgrad}
\end{align}

It is clear that if $\tilde{\mu}_{2g}=\tilde{\mu}_{2s}=0$, all of the expressions from Sec.~\ref{sec:3L} are recovered.  In the limit where $\Delta_e \gg \Omega_m,\sqrt{d \gamma/T_1}$, then $P_2$ can be adiabatically eliminated and the results of Sec.~\ref{sec:3L} are again recovered.  Thus, the extension to a four level system reproduces the results of the three level case in the appropriate limits.

\begin{figure}[h]
\centering
\includegraphics[width=0.95\columnwidth, clip=true]{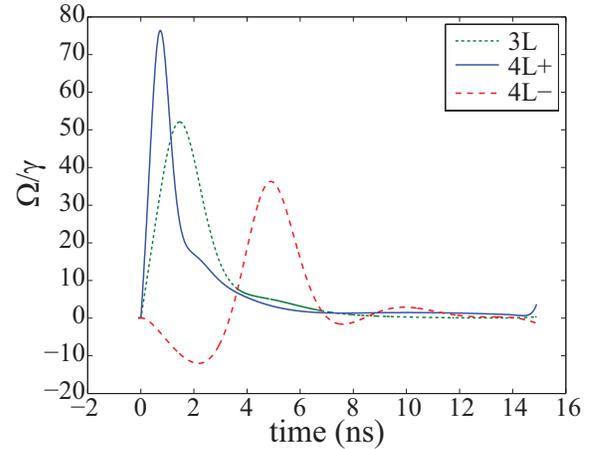}
\caption{(Color online) Optimal control pulses for quantum dot photon storage in a three level system (short-dotted green), the 4L+ four level scenario (solid blue), and the 4L- four level scenario (long-dotted, red).}   \label{fig:fig6}
\end{figure}
To gain some insight into the ramifications of including a fourth level in an intermediate regime of $\Delta_e$, we directly compare the results of the gradient ascent algorithm applied to the three level case to that for two scenarios of the four level case.  For all calculations, we again store a resonant quantum dot photon with $\mathcal{E}_{\textrm{in}}(\tau) = \Theta(\tau)\exp(-\tau/2 T_1)/\sqrt{T_1}$ using a resonant control pulse ($\delta_g=\delta_s=0$).  For the parameters of the storage medium we take $\gamma = 2\pi\times3.035$ MHz, $d=75$, and $\tilde{\mu}_{1g}=\tilde{\mu}_{1s}=1$.  The first four level scenario (4L+) we consider is $\Delta_e=2\pi\times100$~MHz and $\tilde{\mu}_{2g}=\tilde{\mu}_{2s}=1$ while the second scenario (4L-) has the same excited state energy splitting but $\tilde{\mu}_{2g}=-\tilde{\mu}_{2s}=1$.  The optimal control pulses resulting from the gradient ascent optimization are plotted in Fig.~\ref{fig:fig6} for each scenario.  Not only does the inclusion of a fourth level dramatically change the optimal control pulse, but the sign of the relative dipole moment also has a large effect.  More importantly, the computed storage (total) efficiencies are 73.6$\%$, 77.6$\%$, and 43.5$\%$ (63.4$\%$, 65.7$\%$, and 26.3$\%$) for the 3L, 4L+, and 4L- scenarios respectively.  If one naively uses the optimal control for 3L in the 4L+ case, the storage (total) efficiency is only 56.5$\%$ (48.4$\%$), while for the 4L- case 20.8$\%$ (4.8~$\%$) is obtained.  From this analysis it is clear that additional excited states must be taken into account and a careful choosing of those levels is required for any implementation of broadband photon storage where the peak Rabi frequency of the control pulse is comparable to the excited state splitting.

\begin{figure}[h]
\centering
\includegraphics[width=0.95\columnwidth, clip=true]{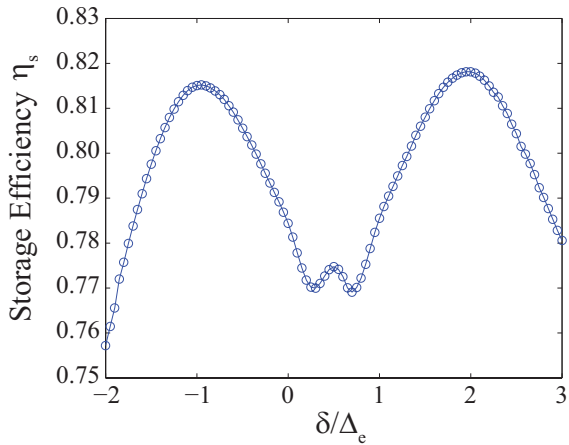}
\caption{(Color online) The storage efficiency after gradient ascent optimization as a function of detuning $\delta=\delta_g=\delta_s$.  The slight asymmetry about $\delta/\Delta_e=0.5$ is due to the finite accuracy of the numerics (see Appendix~\ref{app:Num}).}   \label{fig:fig7}
\end{figure}
Another question one could ask is if there is an optimal choice of the optical detuning $\delta=\delta_g=\delta_s$ in the four level system.  To address this, we again apply the gradient ascent optimization to the 4L+ configuration described above, but we vary the detuning $\delta$ from $-2\Delta_e$ to $3\Delta_e$ in steps of $\Delta_e/20$.  The storage efficiency is plotted in Fig.~\ref{fig:fig7} as a function of $\delta$.  Surprisingly, the efficiencies are smallest near the mid-point between $\ket{1}$ and $\ket{2}$ at $\delta = \Delta_e/2$, while the maxima occur near $-\Delta_e$ and $2\Delta_e$.  Nonetheless, the change in storage efficiency over the entire range is within $\approx5\%$, and there is not much to be gained or lost by changing the detuning at least for this set of $\tilde{\mu}_{\alpha \beta}$.  We have also verified that varying the detuning has a small effect ($\lesssim 5\%$ variation) when using the dipole moments for the relevant $^{87}$Rb transitions.  With these mathematical and numerical tools in hand, we can address storage of a quantum dot photon using the actual atomic levels of $^{87}$Rb.


\section{Storage of a QD-generated Photon in $^{87}$R\lowercase{b}}\label{sec:QD}
Because of the long hyperfine ground-state coherence time and the ability to create cold, dense gases, $^{87}$Rb has proven to be a natural choice for implementation of quantum memory schemes.  Indeed, both the 5$^2$P$_{1/2}$ and 5$^2$P$_{3/2}$ excited state manifolds corresponding to the $D_1$ and $D_2$ optical transitions have been used for quantum memory applications.  Before assigning $^{87}$Rb hyperfine states $\ket{F,m_F}$ to the various states of the four level model, the problem of frequency matching of a QD transition to the $D_2$/$D_1$ transitions at 780/795~nm must be addressed.  The most commonly studied self-assembled QDs are composed of In$_x$Ga$_{1-x}$As islands embedded in a GaAs matrix, which are made to emit light in the 900-1000~nm band.  In such QDs, the so-called ``wetting layer" (a thin quantum well) defines a barrier near 850~nm, below which no confined states exist in the dot.  Therefore, the photons produced by these QDs cannot be stored in $^{87}$Rb, but Cs transitions (852/895 nm) are close to within reach and $^{171}$Yb$^+$ (935 nm) has a transition compatible with QDs~\cite{ref:Waks_PRA09}.  Nonetheless, efficient quantum frequency conversion techniques have been demonstrated using QDs~\cite{ref:Rakher_NPhot_2010,ref:Rakher_PRL_11,ref:Ates_PRL12} such that the single photon produced by a QD could be frequency translated to another wavelength without destroying its quantum characteristics.  Such techniques could bridge the frequency gap between In$_x$Ga$_{1-x}$As QD transitions and those of $^{87}$Rb.  In addition, there are QDs composed of GaAs embedded in AlGaAs and InGaAs QDs embedded in AlGaAs which have been shown to emit in the 780~nm - 795~nm region~\cite{ref:Finley_PRB02,ref:Heyn_APL09,ref:Akopian_APL_2010}.  The optical properties of these QDs are much less well-known compared to standard InGaAs/GaAs QDs and detailed investigations are ongoing.  Thus, either use of quantum frequency conversion techniques or proper material choices permit the study of broadband QD photon storage and retrieval in $^{87}$Rb.  Notably, photons from a QD have been made to interact with $^{87}$Rb atoms in a recent experiment~\cite{ref:Akopian_NPhot2011}.  In this case the atoms acted as a passive medium whose dispersion near the $D_2$ transition was used to reduce the group velocity of the QD photons.  In this present work, we are interested in actively manipulating the atoms to controllably store and retrieve the QD photon.


We proceed by determining which hyperfine states to assign to the four levels of the model.  There are two hyperfine ground-state manifolds in $^{87}$Rb with total angular momentum $F=1$ and $F=2$ separated by 6.835 GHz.  The choice of which ground-states within these manifolds to use is determined by several factors.  Firstly, cross-coupling of the control field and the quantum field can lead to unwanted processes such as four-wave mixing (FWM), which is described in detail in App.~\ref{app:FWM}.  In order to mitigate cross-coupling, it is advantageous to use ground-states that are widely separated in energy and couple to the excited state through perpendicular polarizations.  This becomes especially important at high optical depth and large control fields~\cite{ref:Phillips_PRA11}.  One can choose ground states from the same $F$ manifold, but then the energy splitting is limited to what can be obtained by Zeeman shifting the levels.  For the purposes of QD photon storage, we choose one ground state in $F=1$ and one in $F=2$.  Furthermore, since both states must be coupled to a common excited state, the difference between the $m_F$ values must be between -2 and 2.  This sets the first condition for choosing the ground states.

As discussed in Sec.~\ref{sec:intro}, the long coherence between the hyperfine ground states of $^{87}$Rb make it very attractive as a quantum memory.  This long coherence is usually limited experimentally by magnetic field fluctuations that cause small Zeeman shifts of the $m_F$ levels.  Therefore, choosing $m_F$ levels that are common mode to magnetic field fluctuations for the ground states $\ket{g}$ and $\ket{s}$ is crucial to achieving long storage times.  Notably, it is not possible to find two states in the same $F$ manifold that are common mode to magnetic field fluctuations. However, there are two non-degenerate choices for the ground states from different $F$ manifolds that are common mode to magnetic field fluctuations (to first order) and can be coupled to the same excited state.  These levels are the clock states $\ket{F=1,m_F=0}$ and $\ket{F=2,m_F=0}$ and the states $\ket{F=1,m_F=-1}$ and $\ket{F=2,m_F=1}$ ($\ket{F=1,m_F=1}$ and $\ket{F=2,m_F=-1}$ is a degenerate choice, but is not magnetically-trappable).  For each pair of ground states, there are two optical transitions that are of interest; the $D_2$ transition to the $5^2$P$_{3/2}$ states near 780~nm and the $D_1$ transition to the $5^2$P$_{1/2}$ states near 795~nm.  We will investigate these transitions separately in the following sections.

\subsection{Storage on the $D_2$ Transition}
\begin{figure}[h]
\centering
\includegraphics[width=0.65\columnwidth, clip=true]{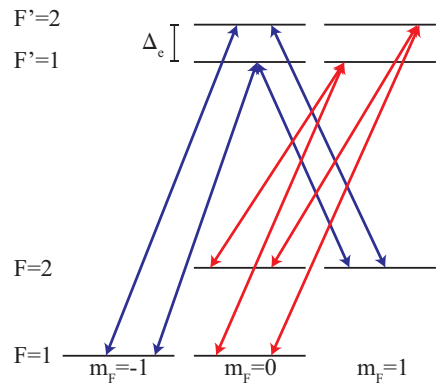}
\caption{Optical transitions of $^{87}$Rb that are eligible for implementation of a quantum memory.}   \label{fig:fig8}
\end{figure}

The excited states of the $D_2$ transition of $^{87}$Rb are composed of four hyperfine manifolds $F'=$ 0, 1, 2, and 3.  Because the two ground states must share a common excited state, only states in $F'=1$ and $F'=2$ are eligible.  For the pair of ground states $\ket{F=1,m_F=-1}$ and $\ket{F=2,m_F=1}$, there are then two possible excited states; $\ket{F'=1,m_F'=0}$ and $\ket{F'=2,m_F'=0}$ as shown in blue in Fig.~\ref{fig:fig8}.

In the following, we always work on resonance with one of the excited states ($\delta_g=\delta_s = 0$) and let the value of $\Delta_e$ change signs.  That is to say, $\ket{g}$ and $\ket{s}$ couple resonantly to $\ket{1}$ and are detuned to $\ket{2}$.  Two choices to assign $\ket{g}$ and two choices for $\ket{1}$ result in four unique configurations for each set of four levels.  To determine which configuration will yield the highest storage efficiency, it is instructive to consider the results of Sec.~\ref{sec:3L}.  Because the storage efficiency depends critically on the optical depth $d$, and the effective optical depth on the $\ket{g}-\ket{1}$ transition is $\tilde{\mu}_{1g}^2 d$, it would seem optimal to choose the configuration where this relative dipole moment is the largest.  A secondary concern is that $\tilde{\mu}_{1s}$ should be large to avoid extremely intense control pulses.  The four relative dipole moments are listed in Table~\ref{tab:D2_1_mu} (see App. A for details), where it can be seen that the largest moment is for the $\ket{F=1,m_F=-1}-\ket{F'=1,m_F'=0}$ transition.
\begin{table}[h]
\centering
\caption{Relative dipole moments $\tilde{\mu}_{\alpha \beta}$ for $^{87}$Rb $D_2$ transitions with ground states $\ket{F=1,m_F=-1}$ and $\ket{F=2,m_F=1}$~\cite{ref:Steck_Rbdata}.}
\begin{tabular}{|c | c| c |}
\hline
& $\ket{F=1,m_F=-1}$ & $\ket{F=2,m_F=1}$ \\ \hline
$\ket{F'=1,m_F'=0}$ & $\sqrt{\frac{5}{12}}$ & $\sqrt{\frac{1}{20}}$ \\ \hline
$\ket{F'=2,m_F'=0}$ & $\sqrt{\frac{1}{12}}$ & $-\sqrt{\frac{1}{4}}$ \\
\hline
\end{tabular}
\label{tab:D2_1_mu}
\end{table}
Taking $\ket{g}=\ket{F=1,m_F=-1}$, $\ket{s} = \ket{F=2,m_F=1}$, $\ket{1} = \ket{F'=1,m_F'=0}$, and $\ket{2} = \ket{F'=2,m_F'=0}$, we can perform gradient ascent to find the efficiencies and control pulse for this configuration when storing a QD single photon with $T_1=1$~ns.  In addition, we have $\Delta_e=2\pi\times156.95$~MHz~\cite{ref:Steck_Rbdata} for the excited state hyperfine splitting and as mentioned previously take both fields to be resonant $\delta_g=\delta_s=0$.  For the sake of comparison, an optical depth $d=75$ is chosen as this has been demonstrated experimentally in ultracold ensembles~\cite{ref:Dudin_PRL_10}.  Note that this corresponds to a standardly-defined optical depth of 150 measured on the cycling transition $\ket{F=2,m_F=\pm2}-\ket{F'=3,m_F'=\pm3}$.  The results of the optimization yield a storage (total) efficiency of 33.6$\%$ (17.3$\%$).  The same calculation was performed for all four configurations of these states (remembering to change the sign of $\Delta_e$ when necessary) and the results for the efficiencies are summarized in Table~\ref{tab:D2_1_eff} along with the peak Rabi frequency of the control pulse $\Omega_m$.
\begin{table}[h]
\centering
\caption{Results of gradient ascent optimization for different configurations of states for $D_2$ using the ground states $\ket{F=1,m_F=-1}$ and $\ket{F=2,m_F=1}$.  The labels $\ket{i,j}$ for states $\ket{g}$ and $\ket{s}$ ($\ket{1}$ and $\ket{2}$) refer to $\ket{F=i,m_F=j}$ ($\ket{F'=i,m_F'=j}$).}
\begin{tabular}{|c|c | c| c |c | c| c|c | c|}
\hline
config & $\ket{g}$ & $\ket{s}$ & $\ket{1}$ & $\ket{2}$ & $\eta_s$ ($\%$) & $\eta_{tot}$ ($\%$)& $\Omega_m$ $(\gamma)$ \\ \hline
1 & $\ket{1,-1}$ & $\ket{2,1}$ & $\ket{1,0}$ & $\ket{2,0}$ & 33.6 & 17.3 & 130.4 \\ \hline
2 & $\ket{1,-1}$ & $\ket{2,1}$ & $\ket{2,0}$ & $\ket{1,0}$ & 30.1 & 12.5 & 58.5 \\ \hline
3 & $\ket{2,1}$ & $\ket{1,-1}$ & $\ket{1,0}$ & $\ket{2,0}$ & 16.6 & 5.4 & 18.0 \\ \hline
4 & $\ket{2,1}$ & $\ket{1,-1}$ & $\ket{2,0}$ & $\ket{1,0}$ & 30.1 & 17.4 & 130.0 \\ \hline
\end{tabular}
\label{tab:D2_1_eff}
\end{table}

\begin{figure}[h]
\centering
\includegraphics[width=0.95\columnwidth, clip=true]{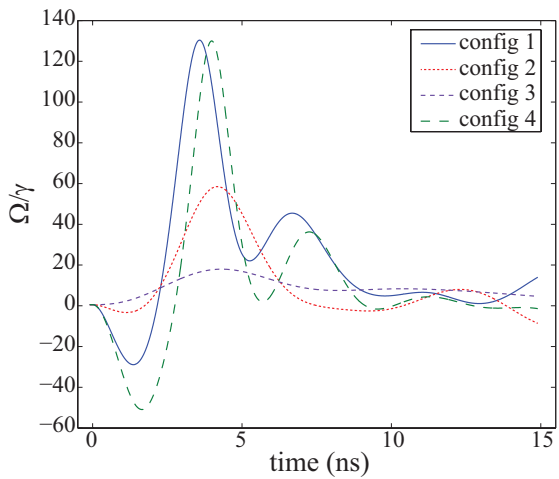}
\caption{(Color online) Optimized control pulses for the configurations detailed in Table~\ref{tab:D2_1_eff}.}   \label{fig:fig9}
\end{figure}

In addition, the optimized control pulses are shown in Fig.~\ref{fig:fig9}.  These results closely align with what is expected; the configurations with the largest relative dipole moments for the $\ket{g}-\ket{1}$ transition have the largest efficiencies.  Nonetheless, these efficiencies are less than what was found in Sec.~\ref{sec:3L} for $d=75$ due to a reduction in the effective optical depth ($\tilde{\mu}_{1g} < 1$) as well as negative effects of the additional excited state $\ket{2}$.  In fact, the relative dipole moments in Table~\ref{tab:D2_1_mu} are such that $\textrm{sgn}(\tilde{\mu}_{1g}\tilde{\mu}_{1s})=-\textrm{sgn}(\tilde{\mu}_{2g}\tilde{\mu}_{2s})$ for all configurations; the same asymmetry that caused a reduction of efficiency in the 4L- scenario studied in Sec.~\ref{sec:4L} due to destructive interference of two-photon pathways.  In order to achieve higher efficiency, such asymmetric configurations should be avoided.

The other choice for the ground states are the clock states $\ket{F=1,m_F=0}$ and $\ket{F=2,m_F=0}$.  As shown in red in Fig.~\ref{fig:fig8}, there are two possible shared excited states; $\ket{F'=1,m_F'=1}$ and $\ket{F'=2,m_F'=1}$ (the pair $\ket{F'=1,m_F'=-1}$ and $\ket{F'=2,m_F'=-1}$ differ only by the photon polarization) again resulting in four unique combinations of relative dipole moments.  The relative dipole moments of these transitions are listed in Table~\ref{tab:D2_2_mu}.
\begin{table}[h]
\centering
\caption{Relative dipole moments $\tilde{\mu}_{\alpha \beta}$ for $^{87}$Rb $D_2$ transitions with ground states $\ket{F=1,m_F=0}$ and $\ket{F=2,m_F=0}$~\cite{ref:Steck_Rbdata}.}
\begin{tabular}{|c | c| c |}
\hline
& $\ket{F=1,m_F=0}$ & $\ket{F=2,m_F=0}$ \\ \hline
$\ket{F'=1,m_F'=1}$ & $\sqrt{\frac{5}{12}}$ & $\sqrt{\frac{1}{60}}$ \\ \hline
$\ket{F'=2,m_F'=1}$ & $\sqrt{\frac{1}{4}}$ & $\sqrt{\frac{1}{4}}$ \\
\hline
\end{tabular}
\label{tab:D2_2_mu}
\end{table}
In contrast to the other set of ground states, these relative dipole moments are symmetric with $\textrm{sgn}(\tilde{\mu}_{1g}\tilde{\mu}_{1s})=\textrm{sgn}(\tilde{\mu}_{2g}\tilde{\mu}_{2s})$ indicating a constructive contribution of the second excited state.  However, because both ground states have $m_F=0$ and couple to the excited states with the same optical polarization, these configurations are sensitive to the effects of four-wave mixing (FWM).  FWM is discussed in detail in App.~\ref{app:FWM} and is shown to be negligible for the parameter regime discussed here.

\begin{figure}[h]
\centering
\includegraphics[width=0.95\columnwidth, clip=true]{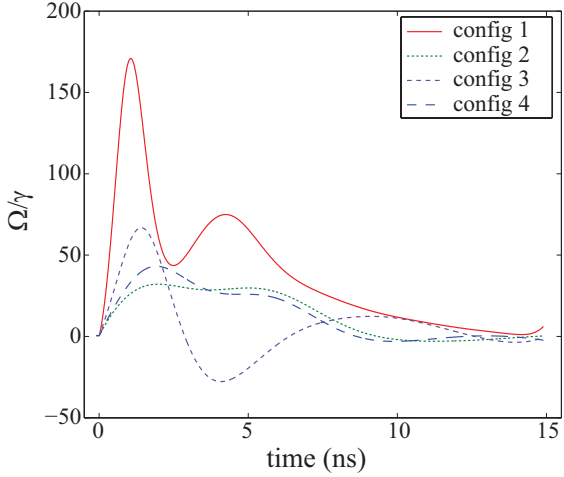}
\caption{(Color online) Optimized control pulses for the configurations detailed in Table~\ref{tab:D2_2_eff}.}   \label{fig:fig10}
\end{figure}

We now perform gradient ascent for each possible configuration using the same set of parameters as before ($d=75$, $\delta_g=\delta_s=0$, and $T_1=1$~ns) and the results are detailed in Table~\ref{tab:D2_2_eff}.  These efficiencies more closely match those found in Sec.~\ref{sec:3L} if the reduction in the optical depth is taken into account.  In fact, the best efficiency for storage on the $D_2$ transition is found using configuration 4 where $\ket{g}=\ket{F=2,m_F=0}$, $\ket{s}=\ket{F=1,m_F=0}$, $\ket{1}=\ket{F'=2,m_F'=1}$, and $\ket{2}=\ket{F'=1,m_F'=1}$.  Storage and retrieval with this configuration yields a storage (total) efficiency of 43.4$\%$ (26.4$\%$).
\begin{table}[h]
\centering
\caption{Results of gradient ascent optimization for different configurations of states for $D_2$ using the ground states $\ket{F=1,m_F=0}$ and $\ket{F=2,m_F=0}$.  The labels $\ket{i,j}$ for states $\ket{g}$ and $\ket{s}$ ($\ket{1}$ and $\ket{2}$) correspond to $\ket{F=i,m_F=j}$ ($\ket{F'=i,m_F'=j}$).}
\begin{tabular}{|c|c | c| c |c | c| c|c | c|}
\hline
config & $\ket{g}$ & $\ket{s}$ & $\ket{1}$ & $\ket{2}$ & $\eta_s$ ($\%$) & $\eta_{tot}$ ($\%$)& $\Omega_m$ $(\gamma)$ \\ \hline
1 & $\ket{1,0}$ & $\ket{2,0}$ & $\ket{1,1}$ & $\ket{2,1}$ & 39.6 & 25.4 & 170.9 \\ \hline
2 & $\ket{1,0}$ & $\ket{2,0}$ & $\ket{2,1}$ & $\ket{1,1}$ & 40.8 & 25.6 & 32.1 \\ \hline
3 & $\ket{2,0}$ & $\ket{1,0}$ & $\ket{1,1}$ & $\ket{2,1}$ & 15.6 & 6.1 & 66.8 \\ \hline
4 & $\ket{2,0}$ & $\ket{1,0}$ & $\ket{2,1}$ & $\ket{1,1}$ & 43.4 & 26.4 & 43.1 \\ \hline
\end{tabular}
\label{tab:D2_2_eff}
\end{table}
The optimized control pulses for each of these configurations are shown in Fig.~\ref{fig:fig10}.  The control pulse for configuration 4 reaches a maximum value of $\Omega_m=43.12\gamma$, which corresponds to a peak power of 12~mW (40~pJ pulse energy) for a Gaussian beam with a 350~$\mu$m waist.  This is roughly three orders of magnitude lower pulse energy than that required to store a photon of similar bandwidth using an off-resonant Raman-based storage scheme~\cite{ref:Reim_NatPhot10} and is achievable using a tunable diode laser and an electro-optic modulator instead of a mode-locked, ultrafast laser.  Thus, reasonable storage and retrieval efficiencies of a quantum dot generated, broadband photon are possible with demonstrated experimental parameters of an ultracold gas of $^{87}$Rb atoms using the appropriate combination of $D_2$ states.  

\subsection{Storage on the $D_1$ Transition}
We now turn our attention to storage and retrieval on the $D_1$ transition of $^{87}$Rb.  Unlike the $D_2$ transition, the excited state is composed of only two manifolds, $F'=1$ and $F'=2$.  In addition, the excited state splitting is quite large $\Delta_e = 2\pi \times 814.5$~MHz so one would anticipate that the state $\ket{2}$ would only play a minor role in the dynamics.  The possibilities for the excited states have the same $F'$ and $m_F'$ values as for the $D_2$ transition, so the level diagram shown in Fig.~\ref{fig:fig8} is also valid.  We proceed to analyze the feasibility of these states by performing gradient ascent for each combination just as was done for $D_2$, taking care to use $\gamma=\pi\times5.75$~MHz and to reference the $\tilde{\mu}_{\alpha \beta}$ correctly~\cite{ref:Steck_Rbdata}.  First, we analyze the possible configurations for the ground states $\ket{F=1,m_F=-1}$ and $\ket{F=2,m_F=1}$.  The relative dipole moments are detailed in Table~\ref{tab:D1_1_mu}.
\begin{table}[h]
\centering
\caption{Relative dipole moments $\tilde{\mu}_{\alpha \beta}$ for $^{87}$Rb $D_1$ transitions with ground states $\ket{F=1,m_F=-1}$ and $\ket{F=2,m_F=1}$~\cite{ref:Steck_Rbdata}.}
\begin{tabular}{|c | c| c |}
\hline
& $\ket{F=1,m_F=-1}$ & $\ket{F=2,m_F=1}$ \\ \hline
$\ket{F'=1,m_F'=0}$ & $-\sqrt{\frac{1}{12}}$ & $\sqrt{\frac{1}{4}}$ \\ \hline
$\ket{F'=2,m_F'=0}$ & $-\sqrt{\frac{1}{12}}$ & $-\sqrt{\frac{1}{4}}$ \\
\hline
\end{tabular}
\label{tab:D1_1_mu}
\end{table}
Notably, each of the $D_1$ relative dipole moments includes a factor of $1/\sqrt{2}$ so they can be directly compared to those for $D_2$ (see Appendix~\ref{app:EoM} for details).  Because of the simplicity of the $\tilde{\mu}_{\alpha \beta}$ in Table~\ref{tab:D1_1_mu}, it is clear that one should obtain the same efficiencies and control pulses if the excited states are interchanged; leaving only two unique configurations.  Gradient ascent optimization is performed for each configuration using the same optical depth and detunings as for $D_2$.  The results are noted in Table~\ref{tab:D1_1_eff}.
\begin{table}[h]
\centering
\caption{Results of gradient ascent optimization for different configurations of states for $D_1$ using the ground states $\ket{F=1,m_F=-1}$ and $\ket{F=2,m_F=1}$.  The labels $\ket{i,j}$ for states $\ket{g}$ and $\ket{s}$ ($\ket{1}$ and $\ket{2}$) correspond to $\ket{F=i,m_F=j}$ ($\ket{F'=i,m_F'=j}$).}
\begin{tabular}{|c|c | c| c |c | c| c|c | c|}
\hline
config & $\ket{g}$ & $\ket{s}$ & $\ket{1}$ & $\ket{2}$ & $\eta_s$ ($\%$) & $\eta_{tot}$ ($\%$)& $\Omega_m$ $(\gamma)$ \\ \hline
1 & $\ket{1,-1}$ & $\ket{2,1}$ & $\ket{1,0}$ & $\ket{2,0}$ & 23.2 & 9.5 & 27.2 \\ \hline
2 & $\ket{2,1}$ & $\ket{1,-1}$ & $\ket{1,0}$ & $\ket{2,0}$ & 44.8 & 28.6 & 77.4 \\ \hline
\end{tabular}
\label{tab:D1_1_eff}
\end{table}
It appears that while the efficiencies for configuration 2 are comparable to those found for the best $D_2$ configuration, the smaller $\tilde{\mu}_{1s}$ for the control transition requires an increase in the peak Rabi frequency $\Omega_m$.  Because the peak intensity (and power) of the control pulse scales as the square of $\Omega_m$, it is experimentally disadvantageous to use this $D_1$ configuration.  In addition, it seems that the sign asymmetry of the $\tilde{\mu}_{\alpha \beta}$ did not play a significant role in reducing the efficiencies, a clear indication that the larger excited state splitting dramatically reduced the effect of an additional excited state.

With that in mind, we now consider the other pair of ground states $\ket{F=1,m_F=0}$ and $\ket{F=2,m_F=0}$.  As stated before, the possible excited states have the same $F'$ and $m_F'$ as for $D_2$ and the relative dipole moments are listed in Table~\ref{tab:D1_2_mu}.  Again, FWM is possible in these configurations but is shown in App.~\ref{app:FWM} to be negligible in the parameter regime considered here.
\begin{table}[h]
\centering
\caption{Relative dipole moments $\tilde{\mu}_{\alpha \beta}$ for $^{87}$Rb $D_1$ transitions with ground states $\ket{F=1,m_F=0}$ and $\ket{F=2,m_F=0}$~\cite{ref:Steck_Rbdata}.}
\begin{tabular}{|c | c| c |}
\hline
& $\ket{F=1,m_F=0}$ & $\ket{F=2,m_F=0}$ \\ \hline
$\ket{F'=1,m_F'=1}$ & $-\sqrt{\frac{1}{12}}$ & $\sqrt{\frac{1}{12}}$ \\ \hline
$\ket{F'=2,m_F'=1}$ & $-\sqrt{\frac{1}{4}}$ & $\sqrt{\frac{1}{4}}$ \\
\hline
\end{tabular}
\label{tab:D1_2_mu}
\end{table}
These $\tilde{\mu}_{\alpha \beta}$ are symmetric under interchange of the ground state, again leaving only two unique configurations.  We perform gradient ascent optimization for each configuration and the results are displayed as configuration 1 and 2 in Table~\ref{tab:D1_2_eff}.
\begin{table}[h]
\centering
\caption{Results of gradient ascent optimization for different configurations of states for $D_1$ using the ground states $\ket{F=1,m_F=0}$ and $\ket{F=2,m_F=0}$.  The labels $\ket{i,j}$ for states $\ket{g}$ and $\ket{s}$ ($\ket{1}$ and $\ket{2}$) correspond to $\ket{F=i,m_F=j}$ ($\ket{F'=i,m_F'=j}$).}
\begin{tabular}{|c|c | c| c |c | c| c|c | c|}
\hline
config & $\ket{g}$ & $\ket{s}$ & $\ket{1}$ & $\ket{2}$ & $\eta_s$ ($\%$) & $\eta_{tot}$ ($\%$)& $\Omega_m$ $(\gamma)$ \\ \hline
1 & $\ket{1,0}$ & $\ket{2,0}$ & $\ket{1,1}$ & $\ket{2,1}$ & 18.5 & 9.0 & 231.8 \\ \hline
2 & $\ket{1,0}$ & $\ket{2,0}$ & $\ket{2,1}$ & $\ket{1,1}$ & 46.0 & 28.9 & 47.4 \\ \hline
3 & $\ket{1,0}$ & $\ket{2,0}$ & $\ket{2,1}$ &  & 45.7 & 28.5 & 45.0 \\ \hline
4 & $\ket{1,0}$ & $\ket{2,0}$ & $\ket{2,1}$ & $\ket{1,1}$ & 45.7 & 28.4 & 45.0 \\ \hline
\end{tabular}
\label{tab:D1_2_eff}
\end{table}
\begin{figure}[h]
\centering
\includegraphics[width=0.95\columnwidth, clip=true]{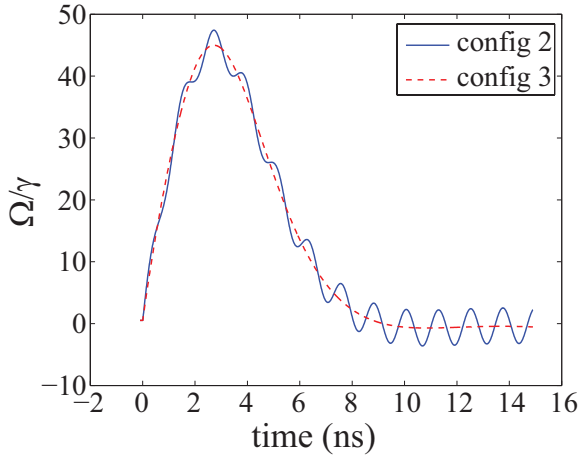}
\caption{(Color online) Optimized control pulses for configurations 2 and 3 detailed in Table~\ref{tab:D1_2_eff}.}   \label{fig:fig11}
\end{figure}
Configuration 2 yields comparable efficiencies to the best configuration for $D_2$, with only a minor increase.  If we perform gradient ascent for this configuration but neglect the additional excited state (configuration 3), we obtain almost identical efficiencies.  Comparison of the two optimal control pulses (Fig.~\ref{fig:fig11}) shows that they are almost exactly the same, except the four level control has a small modulation at the frequency of the excited state splitting.  If we use the three level control on the four level system (configuration 4), we obtain almost exactly the same efficiencies although the control pulse is much simpler to generate.  This analysis simply implies that the large excited state splitting almost completely nullifies the effects of the additional excited state, as anticipated in Sec.~\ref{sec:4L}.  Consequently, the efficiencies we find in this case match well with those predicted by Fig.~\ref{fig:fig3} if $d \tilde{\mu}_{1g}^2 = 18.75$ is used as the effective optical depth of the storage transition.

\section{Imperfect QD Photon Sources}\label{sec:dephase}
Up to this point, our analysis has assumed that the photons emitted by a QD are completely indistinguishable, which is an important property in photon-based quantum information processing~\cite{ref:Knill_Nat_01}.  However, due to their dynamic solid-state environment the photons produced by excitons in quantum dots may not have Fourier transform-limited spectra~\cite{ref:Santori2}.  This leads to an imperfect photon indistinguishability between subsequently emitted photons.  Up to now, the highest measured two photon visibility is $\approx0.97$~\cite{ref:He_Nnano_2013}, corresponding to a linewidth that is $\approx 1.03$ times larger than the transform-limited linewidth $\Delta\omega_{FT} = 1/T_1$.  While it is possible to approach $\Delta\omega_{FT}$ by careful sample selection and resonant excitation, in this section the effect of excess linewidth broadening on the storage efficiency is calculated.

The excess linewidth of quantum dot generated photons can be caused by two distinct physical processes.  The first, sometimes referred to as spectral wandering, is a slow process that causes changes in the carrier frequency of the photons from shot to shot.  This simply means that subsequent photons have slightly different carrier frequencies due to changes in the QD environment on timescales longer than the spontaneous emission lifetime $T_1$.  The second, referred to as pure dephasing, is a perturbation of the QD's energy levels on timescales shorter than $T_1$.  This leads to a time-dependent phase within each shot.  In a time-averaged spectral measurement, these effects can both produce a Lorentzian lineshape with a linewidth greater than the Fourier limit.  Here, we treat these cases separately but show that they cause exactly the same effect on storage and retrieval efficiencies in a quantum memory for a given amount of added linewidth $\Delta\omega_{add}$.

\begin{figure}[h]
\centering
\includegraphics[width=0.95\columnwidth, clip=true]{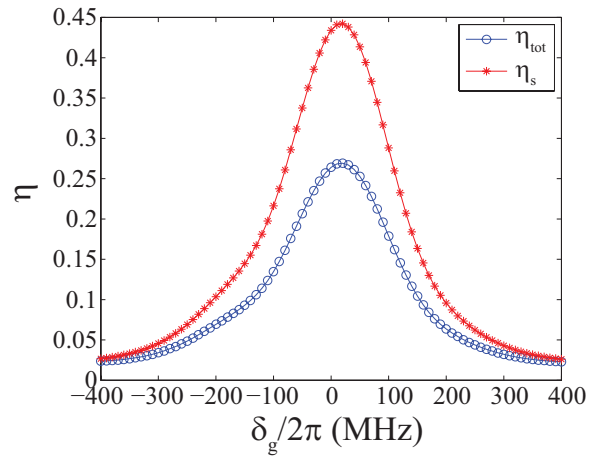}
\caption{(Color online) Storage and total efficiencies for configuration 4 of $D_2$ storage with ground states $\ket{F=2,m_F=0}$ and $\ket{F=1,m_F=0}$ as a function of $\delta_g$.  All other parameters are kept fixed.}   \label{fig:fig12}
\end{figure}

In the case of spectral wandering, the waveform of each photon remains $\mathcal{E}_{\textrm{in}}(\tau) = \Theta(\tau)\exp(-\tau/2 T_1)/\sqrt{T_1}$ but the carrier frequency $\nu_{eg}$ is drawn from a probability distribution $P(\nu_{eg})$.  The time-averaged spectrum of these photons is then an integral over all realizations of $\nu_{eg}$
\begin{equation}
\langle S(\omega) \rangle = \int d\nu_{eg} P(\nu_{eg}) S_{FT}(\omega;\nu_{eg}),
\end{equation}
where $S_{FT}(\omega;\nu_{eg})$ is the transform limited spectrum centered about $\nu_{eg}$.  If $P(\nu_{eg})$ is a Lorentzian distribution with linewidth $\Delta\omega_{add}$, the time-average spectrum is a Lorentzian with linewidth $\Delta\omega_{tot} =\Delta\omega_{FT}+\Delta\omega_{add}$.  The average storage and total memory efficiencies can be calculated similarly by determining how the efficiencies depend on $\delta_g = \nu_{eg}-\omega_{eg}$ and then integrating over all realizations.  We numerically calculate the storage and total efficiencies as a function of $\delta_g$ using the optimal control pulse found for configuration 4 of $D_2$ storage with $\ket{g}=\ket{F=2,m_F=0}$.  All other parameters are kept fixed ($\delta_s=0$, $d=75$, $T_1=1$~ns) and the results are plotted in Fig.~\ref{fig:fig12}.

\begin{figure}[h]
\centering
\includegraphics[width=0.95\columnwidth, clip=true]{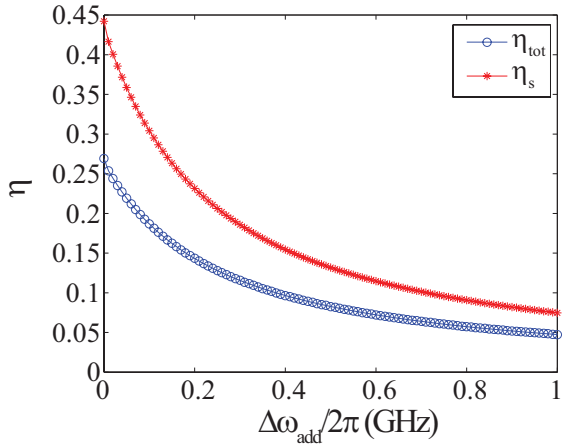}
\caption{(Color online) Storage and total efficiencies for configuration 4 of $D_2$ storage with ground states $\ket{F=2,m_F=0}$ and $\ket{F=1,m_F=0}$ as a function of $\Delta\omega_{add}$ resulting from spectral wandering.}   \label{fig:fig13}
\end{figure}

Using $\eta_s(\delta_g)$ and $\eta_{tot}(\delta_g)$, the efficiencies as a function of $\Delta\omega_{add}$ can be obtained by
\begin{equation}
\eta(\Delta\omega_{add}) = \int d\delta_g \, \eta(\delta_g) P(\delta_g),
\end{equation}
where $P(\delta_g)$ is taken to be a normalized Lorentzian distribution of width $\Delta\omega_{add}$ centered at $\delta_g=0$.  Performing this integration results in the plot shown in Fig.~\ref{fig:fig13}.
Notably, spectral wandering causing a factor of 2 increase in the total linewidth ($\Delta\omega_{add} =\Delta\omega_{FT} $) roughly leads to a factor of 2 reduction in the storage and total efficiencies.  While the initial drop is steep, the total efficiency remains above 5$\%$ even if the total linewidth is more than 7 times the transform limit, showing that the storage process is relatively robust to the effects of spectral wandering.

The effects of fast, pure dephasing can also be calculated numerically.  In this case, the waveform remains at the carrier frequency $\nu_{eg}$ but the slowly varying amplitude is modified to
\begin{equation}
\mathcal{E}_{\textrm{in}}(\tau) = \frac{1}{\sqrt{T_1}}\Theta(\tau)e^{-\tau/2 T_1}e^{-i\phi(\tau)},
\end{equation}
which includes a time-dependent phase $\phi(\tau)$.  In the simplest model of pure dephasing, $\phi(\tau)$ is driven by a Markovian Langevin force $f_{\phi}(\tau)$ characterized by $\langle f_{\phi}(\tau) \rangle = 0$ and $\langle f_{\phi}(\tau) f_{\phi}(\tau') \rangle  = D_{\phi} \delta(\tau-\tau')$ with diffusion constant $D_{\phi}$.  This force will cause phase diffusion, resulting in a Lorentzian time-averaged spectrum with total linewidth $\Delta\omega_{tot} = \Delta\omega_{FT} + D_{\phi}$.  The effect on the storage and total efficiency can be determined by calculating the efficiencies for several trajectories of $\phi(\tau)$ and averaging.
\begin{figure}[h]
\centering
\includegraphics[width=0.95\columnwidth, clip=true]{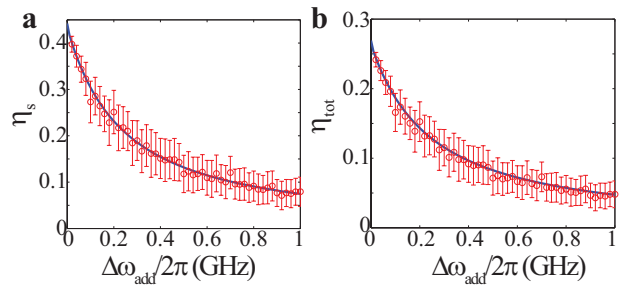}
\caption{(Color online) Average storage (a) and total (b) efficiencies for configuration 4 of $D_2$ storage with ground states $\ket{F=2,m_F=0}$ and $\ket{F=1,m_F=0}$ as a function of $D_{\phi}=\Delta\omega_{add}$ resulting from pure dephasing (red, open circles).  Uncertainties in $\eta$ are given by the standard deviation of the ensemble.  For comparison, results from spectral wandering (Fig.~\ref{fig:fig13}) are shown as blue lines.    }   \label{fig:fig14}
\end{figure}
Using the same configuration as for the spectral wandering investigation, we performed storage and retrieval calculations for 100 phase trajectories for several values of $D_{\phi} = \Delta\omega_{add}$.  The average storage and total efficiencies are plotted as red circles in Fig.~\ref{fig:fig14}, with the uncertainties given by the standard deviation.  For comparison, results from spectral wandering (Fig.~\ref{fig:fig13}) are shown as blue lines.  Evidently, both spectral wandering and pure dephasing lead to the same reduction in efficiency for the same amount of $\Delta\omega_{add}$, indicating that the efficiencies depend only on the time-averaged spectrum and not on the physical mechanism causing an excess linewidth.

%

\section{Ultrahigh Optical Depth}\label{sec:highOD}
A recent experiment by Sparkes \textit{et al.} has demonstrated an optical depth of 1000 in an ultracold $^{87}$Rb gas~\cite{ref:Sparkes_Arxiv_12} using spatial and temporal dark spots.  We therefore investigate how the efficiencies increase for such high optical depth.  If the optical depth is increased to $d$=500 in the optimization using the clock ground states ($\ket{F=1,m_F=0}$ and $\ket{F=2,m_F=0}$) and configuration 4 of $D_2$ storage, the storage (total) efficiency is increased to 51~$\%$ (34~$\%$).  If instead configuration 2 of $D_1$ storage with clock ground states is used, the optimization yields a storage (total) efficiency of 82~$\%$ (76~$\%$).  Notably, the efficiencies are not dramatically increased for $D_2$ storage while $D_1$ storage is much more promising.  There are two factors that contribute to the worse performance of $D_2$; the small excited-state splitting and the large relative dipole element between the storage state $\ket{s}$ and the unwanted excited state $\ket{2}$.  Naively, one would expect an increase in efficiency according to Fig.~\ref{fig:fig3}, but that is only true if the additional excited state doesn't play a large role in the storage process.  However, as the optical depth increases so does the required control pulse amplitude and therefore the coupling to unwanted excited states.  For the case of $D_2$ storage, the extra excited state cannot be neglected and limits the achievable efficiencies even for large optical depth.  On the other hand, for $D_1$ storage of a 1~ns photon at $d=500$ we find very high efficiencies, comparable to the efficiency of 80~$\%$ reported in Ref.~\cite{ref:Sparkes_Arxiv_12} for much lower bandwidth photons.

\section{Conclusion}\label{sec:conc}
In conclusion, we have calculated the efficiency with which a quantum dot generated single photon can be stored and retrieved from an optically-thick $^{87}$Rb ensemble.  Our calculations take into account the multi-level structure of $^{87}$Rb by extending the standard three-level model of an atomic ensemble quantum memory.  Using an optical depth of 150, the storage (total) efficiency can reach 46$\%$ (28$\%$) for a photon resulting from the 1~ns spontaneous excitonic decay in a quantum dot.  Importantly, this storage can be performed using control pulses obtained from a diode laser rather than requiring pulse energies only achievable with ultrafast, mode-locked laser sources.  Increasing the optical depth, for example by using Bose-condensed ensembles, an optical cavity, or advanced trapping techniques can increase the efficiencies to near unity for storage on the $D_1$ transition.  In addition, we have studied the effects of spectral diffusion and pure dephasing of the quantum dot generated photons on the storage efficiency and shown that a factor of 2 increase in the time-averaged photon linewidth roughly leads to a factor of 2 reduction in the efficiency.  Thus, storage and retrieval of single photons from a quantum dot in an $^{87}$Rb ensemble is feasible with demonstrated experimental parameters even in the presence of non-ideal properties of the quantum dot.  Integration of quantum dot sources with atomic ensemble quantum memories may lead to high-speed quantum networks for communication or distributed computation as well as entanglement between collective atomic degrees of freedom and the spin of an electron or hole confined in the quantum dot.
\begin{acknowledgments}

We thank A.~V. Gorshkov, M. Fleischhauer, and K. Srinivasan for helpful discussions.  M.~T.~R. acknowledges support from a Marie Curie International Incoming Fellowship.  This work was also supported by the NCCR Quantum Science and Technology.
\end{acknowledgments}

\appendix

\section{Derivation of the Equations of Motion}\label{app:EoM}
The equations of motion Eq.~\ref{eqn:4LEoME}-\ref{eqn:4LEoMS} (and by extension Eq.~\ref{eqn:3LEoME}-\ref{eqn:3LEoMS}) are derived by considering an ensemble of $N$ motionless, four-level atoms interacting with a quantum field and a classical field closely following the treatment of Ref.~\cite{ref:Gorshkov_PRA2}, whose notation we also adopt for the most part.   The Hamiltonian of this system can be expressed as $\mathcal{H} = \mathcal{H}_{o}+\mathcal{H}_{\textrm{int}} $ where
\begin{align}
\mathcal{H}_{o} &= \sum_i^N \left( E_g \hat{\sigma}_{gg}^i+ E_s \hat{\sigma}_{ss}^i+E_1 \hat{\sigma}_{11}^i+E_2 \hat{\sigma}_{22}^i \right) \nonumber \\
&+\int d\omega\, \hbar \omega \hat{a}^{\dag}_{\omega} \hat{a}_{\omega}
\end{align}
and the interaction between the light fields and the atoms in the dipole approximation is
\begin{equation}
\mathcal{H}_{\textrm{int}} = -\sum_i^N \sum_{\alpha \beta} {\hat{\sigma}}_{\alpha \beta}^i \boldsymbol{\mu}_{\alpha \beta} \cdot \mathbf{\hat{E}}_{\textrm{tot}}(z_i,t).
\label{eq:AppA_Hint}
\end{equation}
Here, the operators $\hat{\sigma}_{\alpha \beta}^i= |\alpha \rangle \langle \beta |$ change the internal state of the $i$th atom from $\ket{\beta}$ to $\ket{\alpha}$ and $\boldsymbol{\mu}_{\alpha \beta}$ is the dipole moment of an atom for the $\ket{\beta}-\ket{\alpha}$ transition.  The total electric field $\mathbf{\hat{E}}_{\textrm{tot}}$ is composed of a classical field $\mathbf{E}_{es}$ and a quantum field $\mathbf{\hat{E}}_{eg}$.  The $+z$-propagating classical field with polarization orientation $\boldsymbol\epsilon_{es}$ can be written as
\begin{equation}
\mathbf{E}_{es}(z,t) = \boldsymbol{\epsilon}_{es} \mathcal{E}_{es}(t-z/c) \cos{\left[ \nu_{es}(t-z/c) \right]},
\end{equation}
where $\mathcal{E}_{es}(t-z/c)$ is the envelope function and $\nu_{es}=\omega_{es}+\delta_s$ is the carrier frequency.  We have assumed that the classical pulse propagates with a group velocity of $c$, which is valid if almost all of the atomic population remains in state $\ket{g}$.  The quantum field is taken to be a sum of modes $\hat{a}_{\omega}$ centered about frequency $\nu_{eg}=\omega_{eg}+\delta_g$ with polarization $\boldsymbol\epsilon_{eg}$ and cross-sectional area $A$
\begin{equation}
\mathbf{\hat{E}}_{eg}(z) = \boldsymbol{\epsilon}_{eg} \sqrt{\frac{\hbar \nu_{eg}}{4\pi c \epsilon_o A}} \int d\omega \hat{a}_{\omega} e^{i \omega z/c}+\textrm{H.c.}
\end{equation}
where H.c. denotes the Hermitian conjugate.  In this treatment, the quantum field only drives the $\ket{g}-\ket{1}$ and $\ket{g}-\ket{2}$ transitions while the classical control field drives the $\ket{s}-\ket{1}$ and $\ket{s}-\ket{2}$ transitions.  As shown in App.~\ref{app:FWM}, cross-coupling of the control field leads to a four-wave mixing process which can reduce the storage efficiency, but it is safe to neglect for the parameter regime considered here.  Applying this assumption and making the rotating wave approximation allows Eq.~\ref{eq:AppA_Hint} to be written as the sum of the interaction with the classical field
\begin{align}
\mathcal{H}_{\textrm{int,c}} &= -\hbar \sum_i^N \Big\{ \Omega_{1s}(t-z_i/c) \hat{\sigma}_{1s}^i e^{-i\nu_{es}(t-z_i/c)} \nonumber \\
& + \Omega_{2s}(t-z_i/c) \hat{\sigma}_{2s}^i e^{-i\nu_{es}(t-z_i/c)}+\textrm{H.c.} \Big\}
\end{align}
and the quantum field
\begin{align}
\mathcal{H}_{\textrm{int,q}} &= -\hbar \sqrt{\frac{L}{2\pi c}} \sum_i^N \int d\omega \Big\{ g_{1g}\hat{a}_{\omega} \hat{\sigma}_{1g}^i e^{i \omega z_i/c} \nonumber \\
&+g_{2g}\hat{a}_{\omega} \hat{\sigma}_{2g}^i e^{i \omega z_i/c} + \textrm{H.c.} \Big\},
\end{align}
where $\Omega_{\alpha\beta}(t-z_i/c) = \boldsymbol{\mu}_{\alpha\beta}\cdot\boldsymbol{\epsilon}_{es} \mathcal{E}_{es}(t-z_i/c)/(2 \hbar)$ are the Rabi frequencies associated with the classical field and $g_{\alpha \beta} =\boldsymbol{\mu}_{\alpha\beta}\cdot\boldsymbol{\epsilon}_{eg} \sqrt{\frac{\nu_{eg}}{2 \hbar \epsilon_o A L}} $ are the couplings to the quantum field.  Note that the Rabi frequencies $\Omega$ are defined differently compared to the standard definition such that $\Omega = \Omega_{std}/2$.

In order to treat the ensemble as a continuous density distribution, we divide the ensemble into thin slices of thickness $L_z$ such that the quantum field can be taken to be constant over this range while also ensuring that the number of atoms in a slice $N_z\gg$1.  Then, we define slowly varying operators
\begin{align}
\hat{\sigma}_{\alpha \alpha}(z,t) &= \frac{1}{N_z}\sum_{i=1}^{N_z} \hat{\sigma}_{\alpha \alpha}^i(t), \\
\hat{\sigma}_{12}(z,t) &= \frac{1}{N_z}\sum_{i=1}^{N_z} \hat{\sigma}_{12}^i(t), \\
\hat{\sigma}_{1s}(z,t) &= \frac{1}{N_z}\sum_{i=1}^{N_z} \hat{\sigma}_{1s}^i(t) e^{-i \nu_{es} (t-z_i/c)}, \\
\hat{\sigma}_{2s}(z,t) &= \frac{1}{N_z}\sum_{i=1}^{N_z} \hat{\sigma}_{2s}^i(t) e^{-i \nu_{es} (t-z_i/c)}, \\
\hat{\sigma}_{1g}(z,t) &= \frac{1}{N_z}\sum_{i=1}^{N_z} \hat{\sigma}_{1g}^i(t) e^{-i \nu_{eg} (t-z_i/c)}, \\
\hat{\sigma}_{2g}(z,t) &= \frac{1}{N_z}\sum_{i=1}^{N_z} \hat{\sigma}_{2g}^i(t) e^{-i \nu_{eg} (t-z_i/c)}, \\
\hat{\sigma}_{sg}(z,t) &= \frac{1}{N_z}\sum_{i=1}^{N_z} \hat{\sigma}_{sg}^i(t) e^{-i (\nu_{eg}-\nu_{es}) (t-z_i/c)}, \\
\hat{\mathcal{E}}(z,t) &= \sqrt{\frac{L}{2\pi c}} e^{i \nu_{eg} (t-z/c)} \int d\omega \, \hat{a}_{\omega}(t) e^{i \omega z/c}.
\end{align}
Using these operators, we can rewrite $\mathcal{H}_o$ as
\begin{align}
\mathcal{H}_o &= \int_0^L dz \, n(z) \Big[ E_g \hat{\sigma}_{gg}(z,t)+E_s \hat{\sigma}_{ss}(z,t) \nonumber \\
&+E_1 \hat{\sigma}_{11}(z,t)+E_2 \hat{\sigma}_{22}(z,t) \Big] + \int d\omega\, \hbar \omega \hat{a}^{\dag}_{\omega} \hat{a}_{\omega}
\end{align}
and the two parts of the interaction as
\begin{align}
\mathcal{H}_{\textrm{int,c}} &= -\hbar \int_0^L dz \, n(z) \Big[\Omega_{1s}(t-z/c)\hat{\sigma}_{1s}(z,t) \nonumber \\
& +\Omega_{2s}(t-z/c)\hat{\sigma}_{2s}(z,t)+\textrm{H.c.}  \Big] \\
\mathcal{H}_{\textrm{int,q}} &= -\hbar \int_0^L dz \, n(z) \Big\{g_{1g} \hat{\mathcal{E}}(z,t) \hat{\sigma}_{1g}(z,t) \nonumber \\
&+ g_{2g} \hat{\mathcal{E}}(z,t) \hat{\sigma}_{2g}(z,t)+ \textrm{H.c.} \Big\},
\end{align}
where $n(z)$ is the linear number density of atoms along the length of the ensemble and we have assumed the cross-sectional area of the field $A$ matches that of the atomic cloud.

The dynamics are determined by finding the Heisenberg equations of motion for the operators.  Using the commutation relations
\begin{align}
\left[ \hat{a}_{\omega}, \hat{a}^{\dag}_{\omega'} \right] &= \delta(\omega-\omega') \\
\left[ \hat{\sigma}_{\alpha \beta}(z,t),\hat{\sigma}_{\lambda \rho}(z',t) \right]&= \frac{\delta(z-z')}{n(z)} \Big\{\delta_{\beta \lambda}\hat{\sigma}_{\alpha\rho}-\delta_{\alpha \rho}\hat{\sigma}_{\lambda\beta} \Big\}
\end{align}
one obtains
\begin{align}
(\partial_t+c\partial_z ) \hat{\mathcal{E}} &= i n L [ g_{g1} \hat{\sigma}_{g1} +g_{g2} \hat{\sigma}_{g2} ] \\
\partial_t \hat{\sigma}_{g1} &= i \delta_g \hat{\sigma}_{g1}+i \Omega_{1s} \hat{\sigma}_{gs} \nonumber \\
&+i \hat{\mathcal{E}} \big[ g_{1g}(\hat{\sigma}_{gg}-\hat{\sigma}_{11})-g_{2g} \hat{\sigma}_{21} \big]\\
\partial_t \hat{\sigma}_{g2} &= i (\delta_g-\Delta_e) \hat{\sigma}_{g2}+i \Omega_{2s} \hat{\sigma}_{gs} \nonumber \\
&+i \hat{\mathcal{E}} \left[g_{2g}(\hat{\sigma}_{gg}-\hat{\sigma}_{22})-g_{1g} \hat{\sigma}_{12} \right]\\
\partial_t \hat{\sigma}_{s1} &= i \delta_s \hat{\sigma}_{s1}+i\Omega_{1s}( \hat{\sigma}_{ss}-\hat{\sigma}_{11})  \nonumber \\
&- i\Omega_{2s}\hat{\sigma}_{21}+i g_{1g} \hat{\mathcal{E}} \hat{\sigma}_{sg}\\
\partial_t \hat{\sigma}_{s2} &= i( \delta_s-\Delta_e) \hat{\sigma}_{s2}+i\Omega_{2s}( \hat{\sigma}_{ss}-\hat{\sigma}_{22})  \nonumber \\
&- i\Omega_{1s}\hat{\sigma}_{12}+i g_{2g} \hat{\mathcal{E}} \hat{\sigma}_{sg}\\
\partial_t \hat{\sigma}_{12} &= -i \Delta_e \hat{\sigma}_{12}+i(\Omega_{2s}\hat{\sigma}_{1s}-\Omega_{s1}\hat{\sigma}_{s2}) \nonumber\\
&+ i(g_{2g} \hat{\mathcal{E}}\hat{\sigma}_{1g}-g_{g1}\hat{\mathcal{E}}^{\dag}\hat{\sigma}_{g2})\\
\partial_t \hat{\sigma}_{gs} &= i(\delta_g-\delta_s)\hat{\sigma}_{gs}+i(\Omega_{s1}\hat{\sigma}_{g1}+\Omega_{s2}\hat{\sigma}_{g2}) \nonumber\\
&- i\hat{\mathcal{E}}(g_{1g}\hat{\sigma}_{1s}+g_{2g}\hat{\sigma}_{2s})\\
\partial_t \hat{\sigma}_{gg} &= i (g_{g1}\hat{\mathcal{E}}^{\dag}\hat{\sigma}_{g1}-g_{1g}\hat{\mathcal{E}}\hat{\sigma}_{1g} \nonumber \\
&+ g_{g2}\hat{\mathcal{E}}^{\dag}\hat{\sigma}_{g2}-g_{2g}\hat{\mathcal{E}}\hat{\sigma}_{2g}) \\
\partial_t \hat{\sigma}_{ss} &= i (\Omega_{s1} \hat{\sigma}_{s1}-\Omega_{1s} \hat{\sigma}_{1s} \nonumber \\
&+\Omega_{s2} \hat{\sigma}_{s2}-\Omega_{2s} \hat{\sigma}_{2s} ) \\
\partial_t \hat{\sigma}_{11} &= i (\Omega_{1s}\hat{\sigma}_{1s}-\Omega_{s1}\hat{\sigma}_{s1}) \nonumber \\
&+ i(g_{1g}\hat{\mathcal{E}}\hat{\sigma}_{1g}-g_{g1}\hat{\mathcal{E}}^{\dag}\hat{\sigma}_{g1}) \\
\partial_t \hat{\sigma}_{22} &= i (\Omega_{2s}\hat{\sigma}_{2s}-\Omega_{s2}\hat{\sigma}_{s2}) \nonumber \\
&+ i(g_{2g}\hat{\mathcal{E}}\hat{\sigma}_{2g}-g_{g2}\hat{\mathcal{E}}^{\dag}\hat{\sigma}_{g2})
\end{align}
where the time and spatial dependencies have been neglected for brevity.

These equations can be reduced considerably by making one simplifying assumption; the quantum field is weak.  The first consequence of this assumption is that almost all atoms remain in $\ket{g}$ for the duration of the dynamics.  Secondly, we keep only terms that are linear in $\hat{\mathcal{E}}$ \cite{ref:Gorshkov_PRA2}.  Under these assumptions, the equations of motion are reduced to
\begin{align}
(\partial_t+c\partial_z ) \hat{\mathcal{E}} &= i n L [ g_{g1} \hat{\sigma}_{g1} +g_{g2} \hat{\sigma}_{g2} ] \\
\partial_t \hat{\sigma}_{g1} &= i \delta_g \hat{\sigma}_{g1}+i\Omega_{1s} \hat{\sigma}_{gs}+i g_{1g} \hat{\mathcal{E}} \\
\partial_t \hat{\sigma}_{g2} &= i (\delta_g-\Delta_e) \hat{\sigma}_{g2}+i\Omega_{2s} \hat{\sigma}_{gs}+i g_{2g} \hat{\mathcal{E}} \\
\partial_t \hat{\sigma}_{gs} &= i (\delta_g-\delta_s)\hat{\sigma}_{gs}+i(\Omega_{s1} \hat{\sigma}_{g1}+\Omega_{s2} \hat{\sigma}_{g2}).
\end{align}
We now introduce the polarization operators $\hat{P}_1(z,t) = \sqrt{N}\hat{\sigma}_{g1}(z,t)$ and $\hat{P}_2(z,t) = \sqrt{N}\hat{\sigma}_{g2}(z,t)$ as well as the spin wave operator $\hat{S}(z,t) = \sqrt{N}\hat{\sigma}_{gs}(z,t)$.  In addition, we move to a new coordinate system $(\tilde{z},\tau)$ where $\tau=t-z/c$ is the time in a co-moving reference frame and $\tilde{z} = (1/N) \int_0^z dz'\, n(z')$ is a dimensionless length.  Inserting these definitions into the equations of motion yields
\begin{align}
\partial_{\tilde{z}} \hat{\mathcal{E}} &= i \sqrt{d_{g1} \gamma} \hat{P}_1 +i \sqrt{d_{g2} \gamma} \hat{P}_2\\
\partial_{\tau}\hat{P}_1 &= (i \delta_g-\gamma) \hat{P}_1 +i\Omega_{1s}\hat{S}+i \sqrt{d_{g1} \gamma} \hat{\mathcal{E}} \\
\partial_{\tau}\hat{P}_2 &= (i\delta_g-i\Delta_e-\gamma) \hat{P}_2 +i\Omega_{2s}\hat{S}+i \sqrt{d_{g2} \gamma} \hat{\mathcal{E}} \\
\partial_{\tau}\hat{S} &= i  (\delta_g-\delta_s)\hat{S} +i\Omega_{1s}^*\hat{P}_1+ +i\Omega_{2s}^*\hat{P}_2,
\end{align}
where a factor of $\sqrt{c/L}$ has been absorbed into $\hat{\mathcal{E}}$ and we introduced the optical depths $d_{\alpha \beta} = g_{\alpha \beta}^2 N L/(\gamma c)$.  We have also assumed that both polarizations decay at the same rate $\gamma$ and that $g_{\alpha \beta}$ is real.

Finally, for notational convenience we compare all transition dipole moments $\mu_{\alpha \beta}$ to that of the two-level cycling transition ($\mu_{\alpha \beta} = \tilde{\mu}_{\alpha \beta}\, \mu_{cyc}$).  In this way, we make the substitutions $\Omega_{\alpha \beta} \rightarrow \tilde{\mu}_{\alpha \beta} \Omega$ and $\sqrt{d_{\alpha \beta}} \rightarrow \tilde{\mu}_{\alpha \beta} \sqrt{d}$ in the equations of motion.  This enables easy comparison between different state configurations and also sets $d = g_{cyc}^2 N L/(\gamma c)$ to where it will be measured experimentally.  It also enables $\Omega$ to be connected to a light intensity through the two-level relation $I/I_s = 2 (\Omega/\gamma)^2$, where $I_s$ is the saturation intensity.  Using the value of $I_s$ for the cycling transition~\cite{ref:Steck_Rbdata}, the peak power can be related to the peak Rabi frequency $\Omega_m$ by
\begin{align}
P_{m} &= \frac{4 \pi^3 \hbar c \gamma}{3 \lambda^3} w_o^2 \left( \frac{\Omega_m}{\gamma} \right)^2 \\
 &= [52.47~\textrm{Wm}^{-2}]w_o^2 (\Omega_m/\gamma)^2 \label{eq:PowerRabi}
\end{align}
where $\lambda$ is the wavelength of the transition and $w_o$ is the $1/e^2$ waist of a Gaussian beam.  The relevant values for the $^{87}$Rb cycling transition $\ket{F=2,m_F=\pm2}-\ket{F'=3,m_F'=\pm3}$ of the $D_2$ line have been inserted to obtain Eq.~\ref{eq:PowerRabi}.  Similarly, the pulse energy can be determined by $U = [52.47$~Wm$^{-2}]w_o^2 \int_0^T d\tau |\Omega(\tau)/\gamma|^2$.  From Ref.~\cite{ref:Steck_Rbdata}, we have
\begin{align}
\mu_{cyc} &= \sqrt{1/2} \langle J=1/2 || e r || J'=3/2 \rangle \\
&= 2.989~e a_o,
\end{align}
where $\langle J=1/2 || e r || J'=3/2 \rangle$ is the reduced dipole moment for the $D_2$ transition.  To put the $D_1$ relative dipole moments $\tilde{\mu}_{\alpha \beta}$ in units of $\mu_{cyc}$, the values in Tables~\ref{tab:D1_1_mu} and \ref{tab:D1_2_mu} have been multiplied by the factor~\cite{ref:Steck_Rbdata}
\begin{align}
r &= \frac{\langle J=1/2 || e r || J'=1/2 \rangle}{\langle J=1/2 || e r || J'=3/2 \rangle} \\
 &= 1/\sqrt{2}.
\end{align}

Further, as discussed in Sec.~\ref{sec:3L} of the main text, the operators can be treated as complex numbers and their associated quantum noise can be neglected because we are interested in computing the expectation values of normally-ordered operators.  Making the substitutions and dropping the operator notation yields the equations of motion as found in Sec.~\ref{sec:3L}-\ref{sec:4L},
\begin{align}
\partial_{\tilde{z}} \mathcal{E} &= i\sqrt{d \gamma}\left[\tilde{\mu}_{1g} P_1 + \tilde{\mu}_{2g} P_2\right] \\
\partial_{\tau}P_1 &= \left(i\delta_g-\gamma \right) P_1+i \tilde{\mu}_{1s} \Omega(\tau) S +i \tilde{\mu}_{1g} \sqrt{d \gamma} \mathcal{E}\\
\partial_{\tau}P_2 &= \left(i\delta_g -i\Delta_e-\gamma\right) P_2 + i \tilde{\mu}_{2s} \Omega(\tau) S \nonumber \\
&+ i \tilde{\mu}_{2g} \sqrt{d \gamma} \mathcal{E} \\
\partial_{\tau}S &= i \left(\delta_g - \delta_s \right) S + i\Omega^*(\tau)\left[ \tilde{\mu}_{1s}P_1 + \tilde{\mu}_{2s}P_2 \right].
\end{align}


\section{Numerical Implementation}\label{app:Num}
Numerical solutions of the equations of motion are obtained using the method of finite differences.  The time-space grid is composed of 9$\times10^6$ (3000 by 3000) points and the domain of $\tau$ is chosen such that the optimized control pulses tend toward 0 at $T$.  This ensures that the drop to $\mathcal{E}_{\textrm{in}}=0$ outside of the domain is smooth.  The gradient ascent algorithm is implemented using a dynamic step size $\lambda$ to guarantee quick convergence.  The step size is determined using an inexact line search such that $\lambda$ is initialized to a large value (1000$\gamma$) at each step of the ascent and the increase of $\eta_{tot}$ and its gradient are calculated at the next step.  If this step does not meet the Wolfe conditions~\cite{ref:Nocedal_numerics}, the step size is reduced geometrically until they are satisfied.  The optimization proceeds until $\eta_{tot}$ has not been increased by more than 0.001$\eta_{tot}$ compared to the average of the three previous values.  This tolerance was estimated by considering the errors resulting from the numerical integration.  Errors in the reported efficiencies are $\approx \pm$1~$\%$ which was determined by examining the variation in efficiencies for perturbations of the time-space grid.  For the parameter set $\delta_g=\delta_s=0$, $d=75$, $\Delta_e = 2\pi\times156.95$~MHz, the gradient ascent optimization took approximately 20 minutes on a standard computer.

The pure dephasing process was simulated by first obtaining $f_{\phi}(\tau)$ on a finite grid using (pseudo-)random numbers drawn from a normal distribution.  Then, $\phi(\tau)$ was calculated using $\phi(\tau_{i+1}) = \phi(\tau_i)+f_{\phi}(\tau_i)$.  Averaging over many trajectories yielded the correct diffusion behavior $\langle[\phi(\tau)-\phi(0)]^2 \rangle = D_{\phi} \tau$.  The extracted $\eta_s$ and $\eta_{tot}$ shown in Fig.~\ref{fig:fig14} were obtained by averaging the results of 100 different phase trajectories for each value of $D_{\phi}$.

\begin{figure}[h]
\centering
\includegraphics[width=0.95\columnwidth, clip=true]{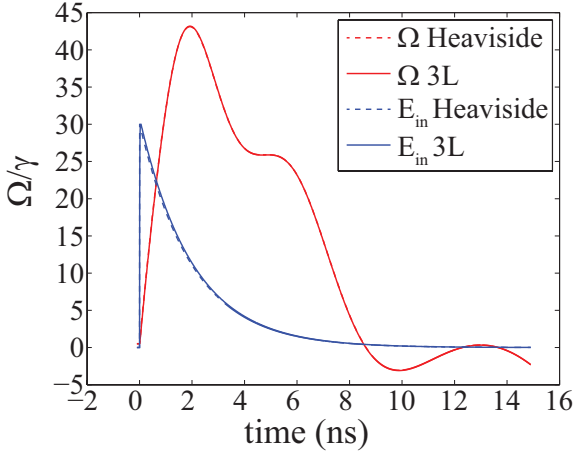}
\caption{(Color online) Optimized control pulses and $\mathcal{E}_{\textrm{in}}$ for the Heaviside model and for the three-level (3L) model ($T_L = 10$~ps) of the QD photon.}   \label{fig:figB_1}
\end{figure}

The slowly varying photon waveform  $\mathcal{E}_{\textrm{in}}(\tau) = \Theta(\tau)\exp(-\tau/2 T_1)/\sqrt{T_1}$ is not physical due to the infinitely sharp rise of $\Theta(\tau)$.  A more realistic model might include the non-zero temporal width of the excitation pulse or fast loading of the QD from another excited state.  For the latter,
\begin{equation}
\mathcal{E}_{\textrm{in}}(\tau) = \Theta(\tau)\sqrt{\frac{e^{-\tau/T_1}-e^{-\tau/T_L} }{T_1- T_L}},
\end{equation}
for a three-level model with instantaneous excitation of an ancillary excited state that loads the QD with rate $1/T_L$.  For comparison, we have implemented this form of $\mathcal{E}_{\textrm{in}}$ with $T_L=10$~ps~\cite{ref:Narvaez_PRB_2006} using configuration 4 of $D_2$ storage with $\ket{g} = \ket{F=2,m_F=0}$.  We obtain roughly the same efficiencies ($\eta_s=43.6\%$, $\eta_{tot}=26.5\%$) and peak control Rabi frequency ($\Omega_m = 43.2\gamma$) as found in Table~\ref{tab:D2_2_eff}.  Both input photons and control pulses are plotted in Fig.~\ref{fig:figB_1}.  Because the curves and efficiencies are extremely similar, we conclude that the infinitely sharp rise of $\Theta(\tau)$ in the simple photon waveform does not dramatically influence the results and is therefore a sufficiently representative choice.

\section{Four-Wave Mixing}\label{app:FWM}
\begin{figure}[h]
\centering
\includegraphics[width=0.65\columnwidth, trim = 0 65 0 0, clip=true]{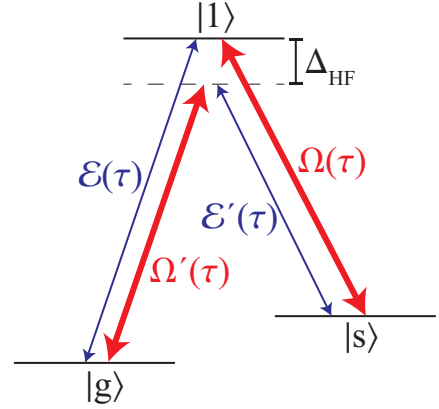}
\caption{Three-level system including cross-coupling for four-wave mixing (FWM). }   \label{fig:figC_1}
\end{figure}

In the preceding analysis, cross-coupling of the control field $\Omega(\tau)$ to the $\ket{g}-\ket{1}$ transition was neglected due to the large ground-state hyperfine splitting of $\Delta_{HF} = 2\pi\times$6.835~GHz.  Because the control field can be quite strong and the optical depth quite large, this cross-coupling can lead to detrimental effects that reduce the storage and retrieval efficiencies in practice~\cite{ref:Hong_PRA09,ref:Phillips_JMO09}.  Of course, a proper choice of ground states and optical polarizations can eliminate cross-coupling completely.  For example, choosing ground states whose $m_F$ values differ like $\ket{g} = \ket{F=1,m_F=-1}$ and $\ket{s} = \ket{F=2,m_F=1}$ and using circularly-polarized light allows cross-coupling to be neglected.  On the other hand, configurations such as $\ket{g} = \ket{F=1,m_F=0}$ and $\ket{s} = \ket{F=2,m_F=0}$ are coupled to the excited state by light of the same polarization.  In this case, the control beam can off-resonantly drive the $\ket{g}-\ket{1}$ transition as shown in Fig.~\ref{fig:figC_1} and coherently generate a Stokes field $\mathcal{E'}$ in a four-wave mixing (FWM) type of process~\cite{ref:Phillips_PRA11} .  The Stokes field can interfere with the spin-wave created from storage of the quantum field $\mathcal{E}$ and lead to reduced storage and retrieval efficiencies.

In Ref.~\cite{ref:Phillips_PRA11}, the Stokes field was taken into account theoretically and was shown to match experimental results quite well.  Following the same approach here, we obtain the Stokes-modified equations of motion for a three-level system
\begin{align}
\partial_{\tilde{z}} \mathcal{E} &= i \tilde{\mu}_{1g}\sqrt{d \gamma}P_1 \label{eqn:3LFWME1}\\
\partial_{\tilde{z}} \mathcal{E'} &= -i \tilde{\mu}_{1g}\tilde{\mu}_{1s}\sqrt{d \gamma} \frac{\Omega}{\Delta_{HF}}S \label{eqn:3LFWME1}\\
\partial_{\tau}P_1 &= \left(i \delta_g-\gamma -2i\delta_{ls} \right) P_1+i \tilde{\mu}_{1s} \Omega S +i \tilde{\mu}_{1g} \sqrt{d \gamma} \mathcal{E} \label{eqn:3LFWMP}\\
\partial_{\tau}S &= i \left(\delta_g - \delta_s-\delta_{ls} \right) S + i \tilde{\mu}_{1s}\Omega^*P_1 \nonumber  \\
&+ i \tilde{\mu}_{1g}\tilde{\mu}_{1s}\sqrt{d \gamma} \frac{\Omega}{\Delta_{HF}} \mathcal{E'^{*}}, \label{eqn:3LFWMS}
\end{align}
where $\mathcal{E'}$ is the Stokes field and the off-resonant interaction $\Omega'(\tau)$ has been adiabatically eliminated, leaving an effective coupling between $\mathcal{E'}$ and $S$.  In addition, this interaction induces time-dependent light shifts of +$\delta_{ls} = \tilde{\mu}_{1g}^2 |\Omega|^2/\Delta_{HF} $ and -$\delta_{ls}$ for states $\ket{1}$ and $\ket{g}$ respectively.  Using these equations of motion with the optimized control field for on-resonance storage of a QD-generated photon with configuration 3 of $D_1$ storage with ground states $\ket{F=1,m_F=0}$ and $\ket{F=2,m_F=0}$ (Fig.~\ref{fig:fig11}), one obtains the same storage and total efficiencies as in Table~\ref{tab:D1_2_eff}.  This result indicates that the effect of FWM in this storage scheme is negligible.

Another way to determine the relative effect of FWM is to consider the ratio of the last two terms in Eq.~\ref{eqn:3LFWMS}.  The ratio of the FWM term to the normal $\Omega^* P_1$ term should roughly scale as $d \gamma^2/\Delta_{HF}^2$, which is approximately 2$\times10^{-5}$ for typical parameters considered here.  Compared to typical parameters found in Ref.~\cite{ref:Phillips_PRA11} where the onset of FWM effects was measured, this ratio is about three orders of magnitude smaller so FWM can be safely neglected.




%

\end{document}